\documentclass{article}
\usepackage{arxiv}
\usepackage[T1]{fontenc}
\usepackage[utf8]{inputenc}
\usepackage{microtype}
\usepackage{amsmath,amssymb,booktabs,tabularx,array}
\usepackage{graphicx}
\usepackage[font=small,labelfont=bf]{caption}
\usepackage[numbers,sort&compress]{natbib}
\usepackage[dvipsnames]{xcolor}
\definecolor{navy}{HTML}{234866}
\usepackage{placeins}
\usepackage{float}
\usepackage[colorlinks=true,linkcolor=navy,citecolor=navy,urlcolor=navy]{hyperref}
\usepackage{xurl}
\usepackage{doi}
\graphicspath{{figures/}}
\newcommand{\papertitle}{Selecting and verifying load-bearing periodic cells for elastic-wave band gaps with numerical acceptance checks}
\newcommand{\paperkeywords}{Periodic lattices, phononic crystals, elastic homogenization, Bloch analysis, numerical verification, engineering software}
\newcommand{\repourl}{https://github.com/nbbllxx0/Designing-Periodic-Cell}

\newcommand{\amm}{AMM}
\newcommand{\E}{E_{100}}
\newcommand{\nuu}{\nu_{100}}
\newcommand{\Cs}{\mathbf C^{*}}
\newcommand{\fig}[1]{Fig.~\ref{#1}}

\newcommand{\statement}[1]{\par\medskip\noindent\textbf{#1}\par\nobreak\smallskip\noindent}
\newcommand{\tablefit}[1]{\begin{tabularx}{\linewidth}{@{}#1@{}}}
\newcommand{\fit}[1]{\begin{tabularx}{\linewidth}{@{}#1@{}}}
\newcommand{\source}[1]{\href{#1}{Source}}
\newcommand{\suppref}[3]{Supplementary #1~#2}

\renewcommand{\suppref}[3]{Supplementary #1~\ref{#3}}
\renewcommand{\shorttitle}{Load-bearing periodic cells with numerical acceptance checks}
\hypersetup{pdftitle={\papertitle},pdfauthor={Shaoliang Yang and Jun Wang},pdfkeywords={\paperkeywords}}

\title{\papertitle}
\author{%
Shaoliang Yang\\
Department of Mechanical Engineering\\
Santa Clara University\\
500 El Camino Real, Santa Clara, CA 95053, USA
\And
Jun Wang\thanks{Corresponding author. Email: \href{mailto:jwang22@scu.edu}{jwang22@scu.edu}.}\\
Department of Mechanical Engineering\\
Santa Clara University\\
500 El Camino Real, Santa Clara, CA 95053, USA
}
\date{November 2026}

\begin{document}
\raggedbottom
\maketitle
\begin{abstract}
A lattice mount must be stiff enough to support equipment and block vibration over a specified frequency range. Fast simulations help rank designs, but their errors can lead to a wrong acceptance decision. We present \amm{} (AgenticMetamaterials), an open-source Python tool that searches repeating cubic cells for targets on stiffness, the fraction of the cell that is solid, and band gaps. A complete band gap is an interval with no elastic-wave propagation in any direction. A separate checker allows for changes as the simulation grid gets finer. For gaps, it also tests for modes at sampled wave directions and wavelengths, so it can miss modes between samples. In 1,500 near-threshold tests of stiffness and solid fraction, the checker accepted 454 of 750 claims supported by the reference and none of the 750 false claims. Using the finest grid estimate alone accepted 42 false claims. In an independent finite-element comparison on ten new cells built from cubes joined by bars, the checker accepted one of two true gap claims and left all 17 false claims undecided. The reference could not classify the remaining claim. The protocol required every reported edge to err on the side of a narrower gap. The checker failed this test when it placed one upper edge too high, even though it left that thin-bar cell's claim undecided. Separating search from acceptance shows which targets the evidence supports and where a calculation needs further refinement.

\end{abstract}

\section{Introduction}

A lattice mount can be stiff enough to support equipment and still transmit vibration in the frequency range of interest. Both properties depend on the geometry of the cell repeated throughout the lattice. The effective elastic tensor describes the static response at scales larger than the cell. The Bloch spectrum describes which elastic waves the infinite lattice can support at each wave vector \citep{laude2020phononic}. A complete band gap is a frequency interval with no propagating mode in any direction. For a mount specified in millimeters, megapascals and kilohertz, the question is whether one cell can meet the stiffness target and keep the requested frequencies inside such a gap.

Fast numerical estimates can help select that cell, but can also place it on the wrong side of a target. For example, a stiffness estimate can change when a thin ligament moves relative to a grid of cubic elements, or voxels. A band plot can miss a mode away from its sampled path, and a finer grid can move an estimated gap edge into the requested interval. Conversely, a search that returns no candidate may have sampled too few geometries. Only a violated analytical bound or consistency condition can rule out the request within the scope of that bound.

We developed \amm{} (AgenticMetamaterials) for cubic periodic cells made from one isotropic solid and void. The tool searches three classes of geometries with one or two continuous parameters per family. It ranks candidates using coarse simulations and a precomputed library of cell properties, called the atlas. A separate checker examines a selected cell against the original targets. It records a certificate when the numerical checks support acceptance, and records why it rejects the candidate or leaves the result unresolved. A request that violates a named bound can be rejected before search. A missed search target can call for a wider search. An unresolved estimate can call for a finer calculation.

We test what each decision means in practice. Solver comparisons measure how the guarded estimates behave against higher-resolution calculations and reference calculations built with the DOLFINx finite-element library \citep{baratta2023dolfinx}. The mount and filter cases carry requirements from physical units through acceptance and mesh export. They also test how a uniform change in the surface affects the result. A budget-matched benchmark compares Bayesian optimization \citep{jones1998efficient} with simpler search methods, and a separate experiment tests the effect of giving one language-model assistant access to the acceptance tools.

The Python library, command line, browser application and Model Context Protocol (MCP) server \citep{mcp2025specification} share the same operations. Supported targets include solid fraction, elastic moduli, Poisson ratio and complete band gaps. The geometry choices include triply periodic minimal surfaces (TPMS), strut lattices and resonators. The workflow can export stereolithography (STL) surface meshes and screen them against specified manufacturing constraints. The study concerns small-strain linear elasticity and numerical acceptance. Manufacturing and in-service performance require separate evidence.

\section{Related software and design methods}

Prior software can generate cells, compute their effective properties or find designs with a requested response. \amm{} adds automatic checks for individual candidates and records the evidence for each decision against the current request. It uses standard numerical methods for homogenization, band calculations, mode counting and refinement.

Geometry generators include LatticeWorks, MSLattice and FLatt Pack, and LatticeWorks includes analysis for selected morphologies \citep{vafaeefar2025latticeworks,alketan2021mslattice,maskery2022flatt}. TPMS Designer generates structures from triply periodic minimal surfaces and provides mechanical analysis \citep{jones2021tpms}. TPMSgen generates minimal-surface cells. Top6Meta also generates strut and other lattice geometries \citep{ecmlab2026top6meta}. \suppref{Table}{S1}{tab:tools} gives the sources and interfaces of both.

For stiffness calculations, the commercial nTop software includes a block that homogenizes a lattice cell under periodic boundary conditions \citep{ntop2026homogenize}. MetamatBench combines metamaterial datasets, learning models, a finite-element evaluator and a visual interface \citep{chen2025metamatbench}. Published homogenization codes and graphics-processing-unit (GPU) solvers provide effective elastic tensors or support inverse design for homogenized properties \citep{dong2019149,andreassen2014how,zhang2023optimized}. FANS (Fourier-accelerated nodal solvers) and fibergen compute effective properties with Fourier-accelerated solvers, and SfePy provides a finite-element homogenization engine \citep{leuschner2018fourieraccelerated,ospald2019fibergen,cimrman2019multiscale}.

Bloch finite elements and dedicated phononic tools calculate dispersion and gaps \citep{krattiger2014bloch,sugawara2026phoxonic}. Lucklum and Vellekoop compared computed gaps for a three-dimensional phononic lattice with transmission measured through printed specimens \citep{lucklum2018bandgap}. Elmadih et al. studied dispersion and transmission numerically for several three-dimensional lattices \citep{elmadih2019multidimensional}. Phononic band-gap materials have been designed by topology optimization \citep{sigmund2003systematic}, locally resonant materials open gaps below the frequencies associated with scattering from the periodic lattice \citep{liu2000locally}, and Hussein et al. review the field\ \citep{hussein2014dynamics}.

Inverse design starts with a target property and searches for a geometry that provides it. Data-driven methods have been used for truss and spinodoid metamaterials \citep{bastek2022inverting,kumar2020inversedesigned}. MetamaterialFinder discovers and analyzes mechanical metamaterials built from simple closed curves \citep{fleisch2024metamaterialfinder}. Single-phase three-dimensional lattices with complete gaps have been designed, printed and measured \citep{warmuth2017single,dalessandro2016modeling}, single-phase resonator cells couple local resonance with Bragg scattering \citep{krushynska2017coupling}, and printed metastructures absorb low-frequency vibration \citep{matlack2016composite}.

We compared the documented functions of 15 packages and research tools. \suppref{Table}{S1}{tab:tools} records the features and points to the source for each entry. Some packages provide components for both homogenization and band calculations. Across the versions inspected, none documented the combination of homogenized stiffness, three-dimensional Bloch gaps, per-result acceptance checks and refusal with bounds. The table records documented features, not measured performance. General-purpose commercial finite-element packages can compute homogenized tensors and Bloch bands through periodic workflows that the user assembles, and a user can apply the per-result checks used here by hand in such a workflow. We did not compare accuracy or run time with commercial packages.

Two numerical practices bear directly on acceptance. Grid-convergence studies extrapolate from three resolutions and report an error band scaled by a safety factor \citep{roache1994perspective}, and the static guard of Section~\ref{sec:methods} follows that pattern. The highest or lowest frequency of a band can occur inside the symmetry-reduced set of wave vectors \citep{harrison2007occurrence,maurin2018probability}, so a gap read from a high-symmetry path can be wrong. Guaranteed bounds are an alternative to extrapolation: boundary conditions on finite samples bracket the apparent elastic moduli \citep{huet1990application}, and methods for self-adjoint operators can place guaranteed or verified bounds on eigenvalues \citep{carstensen2014guaranteed,liu2015framework}. \amm{} uses empirical refinement guards instead, so its certificates are not bounds of that kind.

The design draws on earlier work on evidence-backed release and refusal. A property registry returned a rebuildable catalog cell or a scoped refusal for coupled thermal and mechanical requests \citep{yang2026propertyregistry}. Mechanics benchmarks distinguished recognizing an infeasible brief from rejecting it in a final answer \citep{yang2026language,yang2026calculation}. Work on incomplete topology-optimization specifications addressed when clarification changes the design that can be released \citep{yang2026consequenceguided}. The tool contract, evidence record and replay architecture follow the pattern of AgenticTO \citep{yang2026topologyoptimization} and were re-implemented for periodic cells. The Bloch evaluator, guarded static and gap acceptance, named bounds, error-calibrated atlas, export screen and engineering workflows are new to this work; the static voxel homogenization uses the same voxel discretization and element type as the property registry \citep{yang2026propertyregistry}.

\section{Software design}
\label{sec:software}

\begin{figure}[p]
\centering
\includegraphics[height=0.76\textheight]{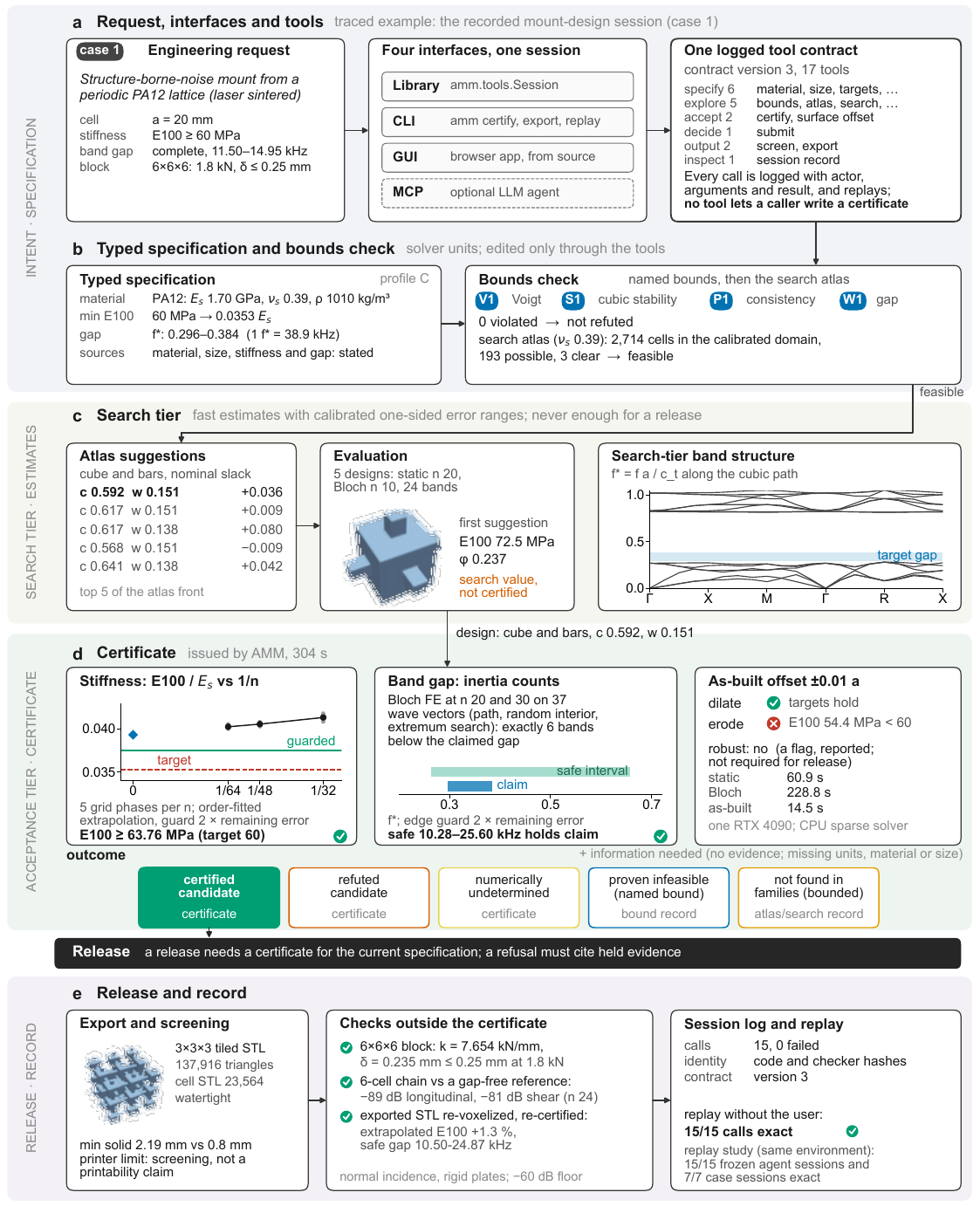}
\caption{The \amm{} workflow. Panels (a)--(d) trace the recorded mount-design session (case~1); panel (e) adds the case-analysis and replay-study records. (a)~Request; four interfaces reach one logged tool contract. (b)~Typed specification, named-bound check and atlas verdict over the cells in the calibrated error domain. (c)~Search-tier estimates, which do not suffice for a release. (d)~Certificate: guarded stiffness, inertia-count gap check with its safe interval, and surface-offset flag. (e)~Export with screening, checks outside the certificate and session replay. Profile C is the full tool profile, in which every refusal must cite evidence. Bound codes: V1, Voigt bound; S1, cubic elastic stability; P1, consistency of the request; W1, gap-request rules. In the atlas verdict a cell is possible when its optimistic estimate meets every target and clear when its pessimistic estimate does. GUI, graphical user interface; MCP, Model Context Protocol; PA12, polyamide 12; CLI, command-line interface; CPU, central processing unit; LLM, large language model; FE, finite elements; STL, stereolithography surface mesh. $n$, voxels per cell edge; $E_s$, Young's modulus of the solid; $f^{*}$, dimensionless frequency of Eq.~\eqref{eq:scale}.}
\label{fig:overview}
\end{figure}

Figure~\ref{fig:overview} traces how a stiffness and band-gap requirement becomes a checked candidate and an export record.

\subsection{Specification, search and decision}

The session stores the base material, cell edge length, allowed geometry families and target values together (\fig{fig:overview}). The base material is defined by Young's modulus $E_s$, Poisson ratio $\nu_s$ and mass density $\rho_s$. The material can come from a preset or be specified explicitly. A user can request bounds on solid fraction, the effective Young's modulus $\E$ and Poisson ratio $\nuu$ along the cubic $[100]$ direction, minima for effective shear and bulk moduli, and a complete gap containing a specified interval. The evaluator also returns the full effective tensor, band structure, directional moduli and the Zener anisotropy ratio $2C_{44}/(C_{11}-C_{12})$, where $C_{11}$, $C_{12}$ and $C_{44}$ are the three independent constants of the cubic effective stiffness tensor. These outputs remain distinct from supported targets.

For example, the mount case specifies polyamide 12 (PA12), a 20~mm cell, a minimum $\E$ of 60.0~MPa and a complete gap over 11.50--14.95~kHz. With the material and length specified, the software can normalize the targets and inspect them before solving. The atlas ranks cells against those targets. Evaluating a suggested cell gives coarse properties and a band diagram that can help the user decide whether to request a certificate. The checker receives the actual candidate and targets. A proposal based on an atlas row or a favorable band plot carries search-tier evidence until the checker runs.

\begin{table}[tbp]
\small
\caption{Outcomes and evidence required by the full tool profile (profile~C). A bound violation concerns the request. A certificate concerns one candidate against the current targets.}
\label{tab:outcomes}
\tablefit{p{0.26\linewidth}X}
\toprule
Outcome & Evidence and scope \\
\midrule
Certified candidate & The checker accepted every target for the specified geometry and material under its numerical assumptions. A gap acceptance rests on an empirical guard (Section~\ref{sec:gapcheck}) with one held-out test (Section~\ref{sec:heldout}). \\
Refuted candidate & A guarded comparison or a geometry check excludes the candidate. \\
Proven infeasible & A named bound or consistency condition excludes the request within its stated scope. \\
Not found in families & The atlas assessment or a budget-limited search returned no qualifying candidate in the declared families. \\
Numerically undetermined & The checker could not decide a target because of convergence, margin or a numerical failure. \\
Information needed & The request lacks information needed to express it in solver units. \\
\bottomrule
\end{tabularx}
\end{table}

Table~\ref{tab:outcomes} separates a failed candidate from a failed request. A minimum modulus above the solid-fraction-scaled Voigt upper bound \citep{voigt1889ueber} can be rejected without searching. If the atlas and the available search budget yield no suitable cell, the outcome is scoped to the families and search used. If a candidate's estimated modulus clears the threshold but its guarded lower value does not, the checker can return an undetermined result. These outcomes lead to different next actions. A bound violation calls for changing a requirement or assumption. A search miss leaves room to widen the design space. Numerical indeterminacy identifies the quantity that needs further resolution.

\subsection{Shared interface and replay}

The Python package exposes 17 operations through version 3 of the tool contract, and results use a versioned outcome schema. All 17 operations belong to the full profile, profile~C. A command-line client, a browser application and an MCP server call the same session implementation.

The browser presents the normalized specification, design preview, estimated properties, guarded values and export controls in one workspace (\suppref{Fig.}{S1}{fig:gui}). The MCP server exposes typed operations that an external client can call directly. The engine accepts the same structured specification without a language model.

A reduced profile, profile~B, exposes 14 operations and omits bounds, certification and the separate surface-offset check. It records a refusal without checking its evidence. We used the reduced profile to test the effect of removing those tools in the agent experiment (Section~\ref{sec:agent}).

The checker alone writes certificates. A certificate records the candidate, normalized targets, material and identifier of the checker build. Editing the requirement therefore prevents release on the strength of a certificate for the prior requirement. For a refusal in the full profile, the cited evidence must also match the current targets and material Poisson ratio. Certificate evidence matches the material, and a search miss must cover the families named in the request.

The session records tool calls and outcomes in order. The session header records the tool code identity, and the certificate records the checker identity. Replay checks numerical results and decisions, with timing excluded from exact comparisons. An STL export can describe an unaccepted candidate as such. The output report carries its evidence status, while a release decision in the full profile requires a matching certificate.

\subsection{Export and manufacturing screening}

The software extracts a triangulated surface from the cell's level set by marching cubes \citep{lorensen1987marching}. It can export one cell or a tiled block in millimeters as an STL file. The accompanying report identifies the result, its evidence and the screening outcome. Screening checks whether the surfaces are watertight, whether the tiled solid is connected and whether solid and void features meet a supplied minimum size. The screen applies a sequence of morphological openings, which remove features too small for the chosen opening size. It reports the smallest size that removes more than 1\% of the solid or void being measured. If that size is below the supplied limit, the screen flags that phase. A thin connector that holds less than 1\% of the solid therefore escapes the screen even when it carries the load path, and the estimate changes with the screening resolution (Section~\ref{sec:cases}). The nominal certificate, the uniform surface-offset flag and this manufacturing screen are reported separately. An exported mesh can also be re-voxelized and checked again, as in Section~\ref{sec:cases}.

\section{Numerical methods}
\label{sec:methods}

The checker takes a candidate and its targets. For a static target, it compares a guarded range of values with the target. A gap claim specifies an entire frequency interval. The checker counts modes inside that interval at sampled wave vectors and also asks whether its estimates of the gap edges leave room for the claim. Both calculations depend on how the grid represents the solid.

\subsection{Material scaling and geometry}

Let $Y$ be the unit cell of edge length $a$. The solvers use $a=E_s=\rho_s=1$ in their normalized variables. Effective moduli scale with $E_s$, and the dimensionless frequency is
\begin{equation}
 f^{*}=\frac{fa}{c_t}, \qquad
 c_t=\sqrt{\frac{E_s}{2(1+\nu_s)\rho_s}},
 \label{eq:scale}
\end{equation}
where $f$ is physical frequency and $c_t$ is the shear-wave speed of the solid. A target in pressure units requires a material modulus. A target in hertz also requires the cell size. A request without the quantities needed for conversion remains incomplete.

The geometry is defined by an implicit surface. Triply periodic minimal surface (TPMS) networks and sheets use a level-set threshold set by the target solid fraction. The choices are gyroid, Schwarz P, diamond and Schoen's I-WP surface (IWP). Simple-cubic, body-centered-cubic and octet strut lattices are parameterized by strut radius and optional nodal-sphere radius. For the resonators, a cube or sphere forms the central mass and bars or ligaments connect neighboring masses. Their core size and connector width vary independently within the family validity rules.

The tool samples points inside each voxel. Near the surface, it estimates the solid coverage of each sub-sample from its signed distance to the surface, using a ramp clipped to the range from zero to full coverage. Voxels with less than 1\% solid coverage are treated as void. The checker examines the largest periodically face-connected component and records the removed fraction and whether the solid spans the cell periodically in all three directions.

A grid phase is a translation of the periodic geometry relative to the voxel grid by the same fraction of a voxel along all three axes. The continuum cell is invariant under this translation, so phase spread measures sensitivity to the placement of the grid. The phases sample translations along the body diagonal only, so their spread can understate the sensitivity to a general translation.

\subsection{Periodic elasticity}

We find the effective stiffness by imposing a unit strain on the cell and computing the local displacement it induces. For a unit macroscopic strain $\boldsymbol\varepsilon_m^0$ in engineering Voigt notation, the periodic displacement fluctuation $\boldsymbol\chi_m$ solves
\begin{equation}
 \int_{Y_s} \boldsymbol\varepsilon(\mathbf v):\mathbf C_s:
 \bigl(\boldsymbol\varepsilon_m^0-\boldsymbol\varepsilon(\boldsymbol\chi_m)\bigr)\,dY=0
 \label{eq:static}
\end{equation}
for every periodic test displacement $\mathbf v$, where $\boldsymbol\varepsilon(\cdot)$ is the symmetric displacement gradient, $Y_s$ denotes the solid and $\mathbf C_s$ its isotropic elasticity tensor. The effective tensor follows by averaging the mutual strain energy over the entire cell,
\begin{equation}
 C^{*}_{mn}=\frac{1}{|Y|}\int_{Y_s}
 \bigl(\boldsymbol\varepsilon_m^0-\boldsymbol\varepsilon(\boldsymbol\chi_m)\bigr):\mathbf C_s:
 \bigl(\boldsymbol\varepsilon_n^0-\boldsymbol\varepsilon(\boldsymbol\chi_n)\bigr)\,dY .
 \label{eq:tensor}
\end{equation}
We reuse one sparse double-precision Cholesky factorization for the six strain load cases. The implementation calls Intel oneMKL's version of the PARDISO direct solver \citep{schenk2004solving} through PyPardiso \citep{haas2025pypardiso}. A pinned node removes the rigid translations. Full voxels use eight-node hexahedra with statically condensed incompatible modes \citep{taylor1976nonconforming}. Partial voxels use finite-cell integration \citep{parvizian2007finite} of the stiffness and consistent mass over the occupied sub-samples.

From $\mathbf S=(\Cs)^{-1}$ the checker computes $\E=1/S_{11}$ and $\nuu=-S_{12}/S_{11}$. A shear target constrains the cube-face shear modulus $G=C_{44}$, not the second cubic shear modulus $(C_{11}-C_{12})/2$. The bulk modulus is $K=(C_{11}+2C_{12})/3$.

\paragraph{Guarded static ranges.}
For a minimum stiffness, the checker compares the lower end of its guarded range with the target. For a maximum, it uses the upper end. To build the range, it tries resolution triplets $(32,48,64)$, $(48,64,80)$ and $(64,80,96)$ in voxels per cell edge. It solves five grid phases at each resolution to account for where the solid falls on the grid. Let $m_1,m_2,m_3$ be the means over phases at grid spacings $h_i=1/n_i$. For a monotone sequence the checker fits an order $p$ in $0.25\le p\le8$ to
\begin{equation}
 m_i=m_\infty+C\,h_i^{p},
 \label{eq:order}
\end{equation}
with limiting value $m_\infty$ and coefficient $C$ (\suppref{Section}{S1}{sec:supp-settings}). The fit screens for monotone, regularly changing sequences. Fitted orders from 0.33 to 6.8 occur in the verification runs, and the fit alone cannot establish that the grids lie in the asymptotic range.

For the guard, we cap the fitted order at $p'=\min(p,1)$. With $r=h_3^{p'}/(h_2^{p'}-h_3^{p'})$, the extrapolated mean and estimated remaining error are
\begin{equation}
 \widehat m=m_3+(m_3-m_2)\,r,\qquad e=|m_3-m_2|\,r .
 \label{eq:guard}
\end{equation}
\paragraph{Small changes without a fitted order.}
When no order fits and the changes are small compared with phase spread or the flat-sequence threshold, the checker uses the finest mean and includes the observed changes in $e$. Unlike the gap guard below, this branch does not test whether the means drift toward the target; Section~\ref{sec:verification} reports the effect of such a test. The guarded range encloses the computed values and $\widehat m\pm2e$ in both branches. For acceptance, each requested quantity must satisfy the tolerances on both the estimated remaining error and the spread across phases. A range that straddles the target leaves the comparison unresolved. The checker can try the next grid triplet when it needs more evidence.

When the fitted order is at least one, its end in the direction of extrapolation, $m_3\pm3|m_3-m_2|\,r$, is the corresponding end of a first-order grid convergence index band with safety factor three \citep{roache1994perspective,celik2008procedure}. The guard is therefore a band of the grid-convergence-index kind with means over grid phases and an order cap. It is used outside the conditions recommended for that index, with refinement ratios of 1.2--1.5 and no check of the asymptotic range: when the cap is reached, the band encloses the limit only if the true order is at least about 0.36.

\paragraph{Early refutation.}
The checker can refute a minimum on $\E$, $G$ or $K$ before convergence if even the largest computed value misses the target. This step assumes that the voxel estimates approach the limit from above. Conforming elements on a fixed geometry would give that property, but the incompatible-mode elements and the smoothed surface remove the guarantee. The calibration cells of Section~\ref{sec:atlas} never gave a search-tier $\E$ or $G$ below the reference, and the 50-cell check of Section~\ref{sec:verification} had no false refutation, but the assumption is not proved, and such a refutation is conditional on it. It is least secure for connectors that span fewer than three voxels, which the feature rule of Section~\ref{sec:gapcheck} flags.

The factor, order cap and phase set come from a study on 11 cells separate from the 50-cell verification set; its scoring also read one cell of that set kept for diagnosis (\suppref{Section}{S1}{sec:supp-settings}). \suppref{Section}{S1}{sec:supp-settings} gives the fit, and \suppref{Table}{S2}{tab:thresholds} lists the solve diagnostics.

\subsection{Bloch bands and gap acceptance}
\label{sec:gapcheck}

Gap acceptance requires both an empty interval in the mode counts and enough margin to the estimated gap edges. The counts test only the requested interval, at sampled wave vectors on the discrete model. The checker estimates the two gap edges separately. To allow for grid refinement, it raises the estimated lower edge and lowers the estimated upper edge. The reported interval between those guarded estimates is called the safe interval. Its endpoints have no separate count test or guarantee for the continuous zone.

Writing a Bloch displacement as $\widetilde{\mathbf u}(\mathbf x)\exp(i\mathbf k\cdot\mathbf x)$ with position $\mathbf x$, wave vector $\mathbf k$ and periodic amplitude $\widetilde{\mathbf u}$ gives the discrete generalized eigenproblem
\begin{equation}
 \mathbf K(\mathbf k)\mathbf U=\omega^2\mathbf M(\mathbf k)\mathbf U .
 \label{eq:bloch}
\end{equation}
Here $\mathbf K(\mathbf k)$ and $\mathbf M(\mathbf k)$ are the Bloch stiffness and mass matrices, $\mathbf U$ the nodal amplitude vector and $\omega$ the angular frequency. The displacement $\mathbf u(\mathbf x)=\widetilde{\mathbf u}(\mathbf x)\exp(i\mathbf k\cdot\mathbf x)$ satisfies the Bloch condition
\begin{equation}
 \mathbf u(\mathbf x+\mathbf a_j)=\exp(i\mathbf k\cdot\mathbf a_j)\,\mathbf u(\mathbf x),\qquad j=1,2,3,
 \label{eq:blochbc}
\end{equation}
for the lattice vectors $\mathbf a_j$ of the cell. During assembly, each element matrix enters the periodic node set with the factor $\exp(i\mathbf k\cdot\mathbf a)$ for a node that wraps across the cell by the lattice translation $\mathbf a$, so $\mathbf K(\mathbf k)$ and $\mathbf M(\mathbf k)$ are Hermitian.

For ordered band frequencies $f_j(\mathbf k)$, the continuous-zone gap above band $j$ lies between $\max_{\mathbf k} f_j(\mathbf k)$ and $\min_{\mathbf k} f_{j+1}(\mathbf k)$ if the latter exceeds the former. The evaluator estimates bands on a high-symmetry path. The checker samples that path and points inside the zone. It tests the requested interval by counting modes directly.

\paragraph{Counting modes in the requested interval.}
At a given wave vector, the checker counts the modes below each claim endpoint. An increase between the two counts signals a mode in the tested interval. The same count must hold at both endpoints and across all sampled wave vectors. For a claim from $f_{\mathrm{lo}}$ to $f_{\mathrm{hi}}$, the count takes the form
\begin{equation}
 N(\mathbf k,\sigma)=\nu_-\bigl(\mathbf K(\mathbf k)-\sigma\mathbf M(\mathbf k)\bigr),
 \label{eq:count}
\end{equation}
the number of negative eigenvalues of the shifted matrix, at the shifts $\sigma_{\mathrm{lo}}$ and $\sigma_{\mathrm{hi}}$ of the two endpoints. In normalized units, a dimensionless frequency $f^{*}$ corresponds to a shift $\sigma=(2\pi c_t f^{*})^2$ with the normalized solid shear speed.

The claim passes the counts when $N(\mathbf k,\sigma_{\mathrm{lo}})=N(\mathbf k,\sigma_{\mathrm{hi}})=j$ at every sampled $\mathbf k$, which places the claimed interval between bands $j$ and $j+1$ at the tested points. The count uses a numerical factorization. Its diagnostics check for gross failures, and rounding can still put a mode on the wrong side of an endpoint (\suppref{Section}{S1}{sec:supp-settings}).

\paragraph{Wave-vector sampling.}
The checker counts modes at $n=20$ and 30 voxels per cell edge. The sampled set has 37 wave vectors: 21 on the path $\Gamma$--X--M--$\Gamma$--R--X, ten random interior points and six points from a coarse search for the extrema of the two bounding bands (\suppref{Section}{S1}{sec:supp-settings}). The path omits the M--R leg, which the dense scan of Section~\ref{sec:verification} covers.

Interior samples lie in the wedge $0\le k_z\le k_y\le k_x\le\pi$, which assumes the full cubic symmetry of the discrete model. A grid shifted by a quarter voxel breaks some mirror symmetries, so the finest counted grid is checked at two further phases (0.25 and 0.5 voxel) as a sensitivity test rather than a sample of the full zone. A third resolution ($n=40$) estimates the edges at the extreme wave vectors from the finest counted grid. This reuse of wave vectors and the limited sampling can make the estimated gap too wide. The edge guard has no allowance for extrema between sampled wave vectors. In the dense scan one band exceeded its sampled estimate by 0.30\%. The default gap check stops there. A gap left undetermined can be checked at finer resolutions through the checker's resolution arguments, as for the filters of Section~\ref{sec:filter}.

\paragraph{Gap-edge estimates.}
The edge guard fits the three estimates of each edge as in Eqs.~\eqref{eq:order} and \eqref{eq:guard}, with the order capped at two. When the cap is reached, the guarded edge encloses the limit only if the true order of the edge is at least about 0.80 on the $n=20$, 30, 40 ladder and 0.86 on $n=30$, 40, 60. For the lower edge, the guard uses the highest of the computed values and the extrapolation shifted inward. For the upper edge, it uses the lowest. With $F_{\mathrm{lo}}$ and $F_{\mathrm{hi}}$ the sets of resolution and phase values of the lower and upper edge, the safe edges are
\begin{equation}
 s_{\mathrm{lo}}=\max\bigl(F_{\mathrm{lo}}\cup\{\widehat f_{\mathrm{lo}}+2e_{\mathrm{lo}}\}\bigr),\qquad
 s_{\mathrm{hi}}=\min\bigl(F_{\mathrm{hi}}\cup\{\widehat f_{\mathrm{hi}}-2e_{\mathrm{hi}}\}\bigr),
 \label{eq:safe}
\end{equation}
and the claim must satisfy $s_{\mathrm{lo}}\le f_{\mathrm{lo}}$ and $f_{\mathrm{hi}}\le s_{\mathrm{hi}}$. Thus the safe interval must contain the whole claim. A failed count alone leaves the gap undetermined because a mode inside the interval on a finite grid need not persist under refinement.

\paragraph{When the checker withholds acceptance.}
When an edge moves toward the claim as the grid gets finer, it must pass the order fit. This applies to a rising lower edge or a falling upper edge. If no order fits between 0.25 and 8, the checker reports a safe value only when both changes lie within twice the phase spread and the edge stays put or moves away from the claim. In that case the computed and phase values bound the edge and the phase spread is its error. \suppref{Table}{S8}{tab:rules} lists the rule changes and the certificates they withdrew.

\paragraph{Resolving thin features.}
Both the static and the gap checks require at least three voxels across the load-path feature at the finest resolution. We define this feature by the diameter of the largest ball that passes through the solid along each axis. This is the bar width of a resonator or the strut diameter of a strut lattice. For a minimal-surface cell, we compute the bottleneck on a $128^3$ grid. Below the threshold, the static checker tries a finer triplet. Any certificate that the threshold still prevents is reported as undetermined. We chose the three-voxel threshold by judgment.

\paragraph{Evidence for the guard settings.}
The static order cap of one and the five-phase set were chosen on 11 cells separate from the 50-cell verification set, and the same study checked the static bounds with the factor of two (\suppref{Section}{S1}{sec:supp-settings}). The gap factor of two was calibrated on one published case \citep{lucklum2018bandgap} with a first-order form of the guard that used only the two counted resolutions, and the gap order cap of two has no held-out calibration. Section~\ref{sec:verification} reports the current guard on that cell.

For gaps, the six finite-cell and eight DOLFINx comparison cells also informed the rule changes. The original guard had graded as sound on these comparisons, although it could read a drift toward the claim as noise. A held-out test on ten further cube-and-bar cells certified no false claim but found one unsound safe edge, on a cell below the feature rule (Section~\ref{sec:heldout}). The counts establish a property of the discretized model at the sampled wave vectors. An unsampled wave vector can still admit a mode inside the interval \citep{harrison2007occurrence,maurin2018probability}. Table~\ref{tab:assumptions} lists the assumptions behind a certificate and where each is checked.

\begin{table}[tbp]
\small
\caption{Assumptions behind a certificate. Run time: checked by the checker for every certificate. Section~\ref{sec:verification}: tested on the verification cells, not at run time.}
\label{tab:assumptions}
\tablefit{p{0.42\linewidth}X}
\toprule
Assumption & Where it is checked \\
\midrule
Phase means are monotone and an order in 0.25--8 fits them & Run time; otherwise the phase-noise branch or abstention \\
The three resolutions lie in the asymptotic range & Assumed; not checked. The order window tests monotonicity only, and fitted orders of 0.33--6.8 occur \\
A gap edge that moves toward the claim follows a fitted order & Run time; otherwise the edge has no safe value \\
Phase spread measures sensitivity to grid placement & Run time: five static phases, two further Bloch phases, all translations along the body diagonal \\
Static order cap one, five phases and factor two & Study on 11 cells separate from the 50-cell set (\suppref{Section}{S1}{sec:supp-settings}) \\
Gap safety factor two & Calibrated on one published gap with an earlier guard; released guard developed on the comparison cells. Held-out test (Section~\ref{sec:heldout}): no false claim certified, one unsound safe edge \\
Gap order cap two & Not calibrated; cap one would change one of 46 benchmark gap verdicts and no comparison, engineering or held-out verdict (Section~\ref{sec:verification}) \\
Three voxels across the load-path feature resolve it & Run time; threshold set by judgment. The unsound held-out safe edge came from a cell below it \\
Sub-sampled geometry is resolved at the finest level & Three engineering certificates; exact-geometry references of two (Sections~\ref{sec:verification}, \ref{sec:cases}). On the finer gap ladder the sub-sample grid is fixed, so $e$ omits geometry sampling \\
Voxel stiffness approaches the limit from above, used for early refutation & Calibration and 50-cell sets; not proved \\
The 37 sampled wave vectors locate the band extrema & Dense scan of twelve gaps at $n=20$ (Section~\ref{sec:verification}): one band exceeded its sampled estimate by 0.30\%; no sampling term \\
The discrete model has full cubic symmetry, used for wedge sampling & Partly, through the two further Bloch phases \\
Counts of the computed factorization equal those of the discrete model & Run-time diagnostics: even counts, no perturbed pivots, three acoustic modes; no margin to the shift \\
Linear elasticity of one isotropic solid with nominal properties & Not checked \\
\bottomrule
\end{tabularx}
\end{table}

\subsection{Analytical bounds and surface offsets}

The Voigt bound $\Cs\preceq\phi\mathbf C_s$ compares quadratic forms for a solid fraction $\phi$ in a solid--void cell. If $\phi$ has a specified upper bound $\phi_{\max}$, a request for $\E>\phi_{\max}E_s$ is infeasible within the model. Analogous upper bounds apply to the shear and bulk moduli. Cubic elastic stability confines $\nuu$ to $(-1,1/2)$. The acoustic branches at zero frequency exclude a complete gap containing zero. Reversed intervals or nonphysical solid fractions also trigger consistency checks. Sharper bounds, such as bounds of the Hashin--Shtrikman type \citep{hashin1963variational}, are not implemented, so a request that a sharper bound would exclude can pass the Voigt check and is not reported as infeasible. These bounds catch gross static violations and malformed gap requests only. No bound couples stiffness to a gap, so a realistic request that no family meets ends as not found in families rather than proven infeasible.

For an as-built flag, the checker erodes and dilates the nominal surface by $0.01a$. The offset scales with the cell edge, 0.2~mm for the 20~mm mount cell and 0.1~mm for the 10~mm filter cells, not with a process tolerance. It repeats static checks at a coarse resolution for the stiffness and Poisson-ratio targets and repeats a smaller set of Bloch counts for a gap target. Failure of either surface offset sets the flag and preserves the nominal decision. The test models a uniform change in surface position and provides no estimate of manufacturing variability. Because it runs at the coarsest static resolution and on a reduced set of wave vectors, it is a screen, not evidence of robustness.

\subsection{Atlas and search ranking}
\label{sec:atlas}

The search tier uses $n=20$ for static homogenization and $n=10$ for Bloch bands, with a dense batched eigenproblem on a GPU when available.

The atlas covers 2,722 designs at each of three solid Poisson ratios, 2,710 of them valid; 2,714 of the rows at each ratio, valid or not, lie inside the calibrated error domain. It stores coarse properties and one-sided error ranges fitted by family against 73 reference cells. Against these cells the search tier never underestimated $\E$ or $G$, but it overstated $\E$ by more than 5\% in 32 cells and $G$ in 50, by up to 73\% and 111\%. In 14 of the 18 cells with a gap it placed the upper gap edge more than 5\% of the mid-gap frequency too high; over all 18 cells the median was 23\%. These are the errors that let a single coarse check accept a candidate that misses a minimum stiffness or an interval near the upper edge. The ranges are widened for thin features.

To rank a candidate, the search measures its distance from each target on a normalized scale. The smallest of these distances, called slacks, sets the candidate score. The atlas can rank by nominal slack or apply median, optimistic or pessimistic error corrections. An atlas verdict uses these error ranges and interpolation between sampled designs. It supplies an estimate within the families, while a proof of infeasibility requires a violated bound.

Bayesian optimization \citep{jones1998efficient} fits a Gaussian process for each selected geometry kind and mode. With atlas seeding, it ranks these search arms by their best atlas score and keeps the three highest. At each step it uses batch log expected improvement \citep{ament2023unexpected} to choose among 1,024 Sobol \citep{sobol1967distribution} candidates, with a batch size of eight. The full search uses the median-corrected slack as the objective. The higher-resolution certificate is run on selected designs after search. Separating these steps lets the search use inexpensive estimates while the acceptance decision remains tied to the requested quantities and the numerical evidence.

\section{Solver and certificate verification}
\label{sec:verification}

We first compare solvers on the same discrete model. Finer-grid runs on the voxel model then test the effect of refinement. Comparisons with finite-element models of the exact geometry also expose errors in the voxel representation. Finally, we compare against one published gap and test the released gap guard on ten cells that did not inform it.

For a reference value $R$ with estimated refinement uncertainty $u$, we grade a guarded bound against the band $R\pm2u$. A bound is sound if it clears the band on the conservative side, marginal if it falls within the band and unsound if it crosses the far side. For a gap, conservative means raising the lower edge or lowering the upper edge. An edge can be sound and still give up much of the reference gap, so we also report how far the guarded edges lie inside it.

The initial comparison studies issued certificates under the original rules. The revised rules withhold acceptance if the load-bearing feature is too thin for the grid or if an edge drifts toward the requested interval without a fitted refinement order (Section~\ref{sec:gapcheck}). A subsequent correction uses the actual grid spacings in the order test. The revised rules with this correction are the released rules. We report decisions under those rules unless stated otherwise, reusing stored solves where needed. \suppref{Table}{S8}{tab:rules} lists the changes and the builds that used them.

\subsection{Static checks}

\begin{figure}[!htbp]
\centering
\includegraphics[width=0.95\linewidth]{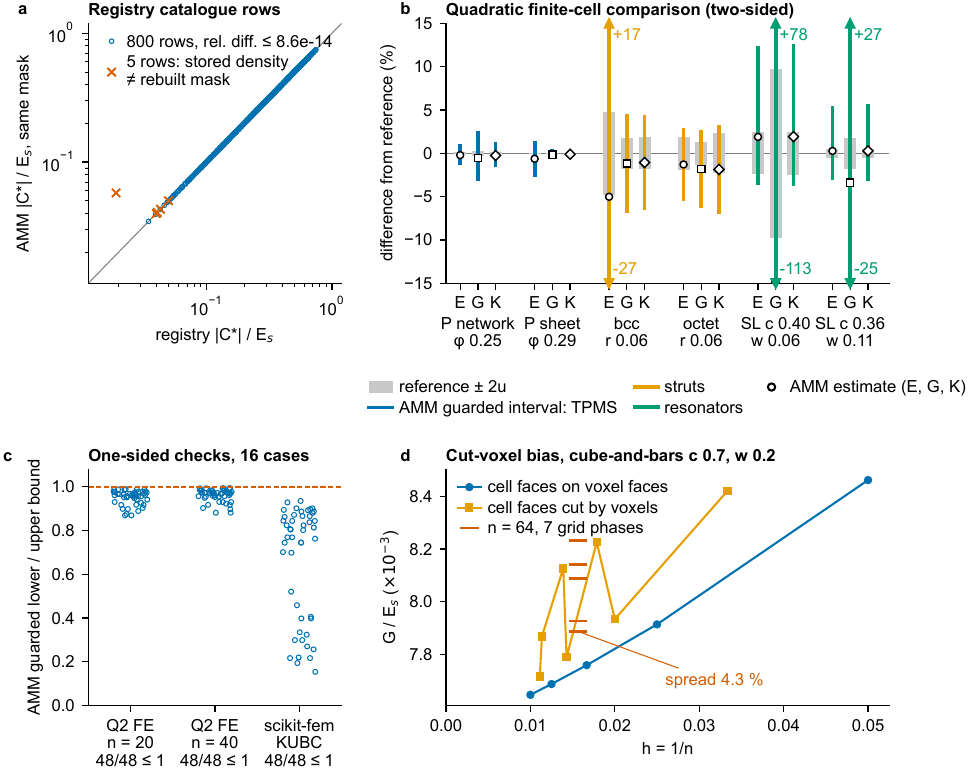}
\caption{Static verification. (a) Tensor-norm parity against the registry on a shared voxel model. Crosses denote discrepancies in stored versus rebuilt density. (b) Guarded intervals (bars) and extrapolations (open markers) for effective Young's modulus $E$, shear modulus $G$ and bulk modulus $K$ in the six-cell quadratic finite-cell comparison. The vertical coordinate is the percentage difference from the reference. Gray denotes reference uncertainty $R\pm2u$. $R$, reference value; $u$, its numerical uncertainty; TPMS, triply periodic minimal surface. Cells: P, Schwarz P network or sheet with solid fraction $\phi$; bcc (body-centered cubic) and octet strut lattices with strut radius $r$; SL, sphere and ligaments with core radius $c$ and ligament diameter $w$, in cell units. (c) Ratios of guarded lower bounds to upper comparisons from quadratic finite elements (Q2) or kinematic uniform boundary conditions (KUBC) \citep{huet1990application}. Values at or below one pass. (d) Shear modulus for a cube-and-bar cell on aligned and cut voxel grids, against the grid spacing $h=1/n$ for $n$ voxels per cell edge.}
\label{fig:static}
\end{figure}

\paragraph{Solver agreement and one-sided comparisons.}
On identical voxel masks and elements, 800 of 805 load-bearing entries from the property registry \citep{yang2026propertyregistry} reproduced the stored tensor norm to a relative difference of at most $8.61\times10^{-14}$ (\fig{fig:static}a). The other five entries had a stored density different from that of the rebuilt mask, and all five were already flagged as infeasible in the registry. The registry tensors come from voxel homogenization of the same masks with the same element type, so this is a parity check of the shared discretization, not a test of accuracy.

A separately written quadratic finite-element implementation supplied one-sided upper comparisons for $\E$, $G$ and $K$ in 16 cube-and-bar cases (10 geometries, 13 cell--material combinations) at two resolutions. All 96 guarded lower values lay below the corresponding upper values. An assembly using scikit-fem \citep{gustafsson2020scikitfem} gave 48 of 48 consistent comparisons with upper bounds obtained from prescribed affine boundary displacements. The comparisons share the geometry evaluator.

\paragraph{Refinement on 50 cells.}
We compared the checker with reference brackets computed by \amm{} on 50 cells. The reference used phase means up to $n=96$ for static quantities and gap edges up to $n=50$. At each cell's finest reference level, which matched or exceeded the finest certificate level, we bracketed the quantity using the finest value and two extrapolations, one with the fitted order and one at first order. A claim is true if it holds over the whole bracket, false if it fails over the whole bracket and undetermined otherwise. These references share \amm{}'s discretization, so they test the guard against refinement, not against the continuum.

Under the final classification rule, the checker made zero false certifications and zero false refutations. All claims held in the reference for 41 cells. The checker certified 35 of those cells and left six undetermined. The final classification criterion was redefined after results existed, so these counts are development evidence for the final rule on this set; the quadratic finite-cell and DOLFINx comparisons below are separate tests.

The same cells show what the checker adds to the error-corrected search tier. Their 263 claims were set 5\% inside a first-pass reference, and four gap claims are false. We evaluated every claim at the search tier and applied the calibrated error ranges of Section~\ref{sec:atlas}. In only one of the 41 cells with positive reference truth did every claim clear the pessimistic estimate, the rule by which the atlas calls a cell feasible. The other 40 cells stayed between the pessimistic and optimistic estimates, and the checker certified 34 of them. Neither tier accepted a false claim: the four false gap claims stayed between the two estimates at the search tier and were left undetermined by the checker. The error ranges were fitted partly on these cells, which favors the search tier in this comparison.

\paragraph{Decisions close to a static target.}
We used the 50 cells to test claims near a decision threshold. For $\E$, $G$ and solid fraction, we moved each threshold 0.5, 1, 2, 3 or 5\% beyond or inside the reference bracket. That gave 750 false and 750 true claims to evaluate against the stored guarded bounds. The checker certified none of the false claims and 454 of the true ones, from 57 of 150 at 0.5\% to 126 of 150 at 5\%. A rule that accepts a claim when the finest computed phase mean meets it, without a guard, accepted 42 false claims, 24 of them at 0.5\%.

We also applied two grid convergence index (GCI) bands \citep{roache1994perspective,celik2008procedure} to the same phase means. Using the fitted order and a factor of 1.25 accepted 3 false and 726 true claims. For the 140 claims with no fitted order, that rule used first order and a factor of 3. Applying first order and a factor of 3 to every claim accepted no false claims and 439 true ones. In the direction of extrapolation, that band's end matches the checker's when the fitted order is at least one (Section~\ref{sec:methods}).

\suppref{Section}{S6}{sec:supp-records} reports which claims differed between the checker and the first-order band, the effect of a stricter static drift rule, and the separate gap-claim comparison.

On 29 of the 50 cells the reference's finest level is the certificate's own finest level, which limits what a false claim can test there. On the other 21 cells, whose reference is strictly finer, the checker certified none of 315 false claims and 215 of 315 true ones, the unguarded rule accepted seven false claims and both grid convergence index bands none.

In a cube-and-bar example, moving the geometry relative to the grid changed the inferred shear modulus by 4.3\% at $n=64$ (\fig{fig:static}d). A fixed grid can therefore hide an error that matters near a shear target.

\paragraph{Two-sided finite-cell comparison.}
We selected six cells before seeing any comparison results: two cells of the 50-cell check and four previously certified designs, three of them with small-margin claims. The reference was a quadratic finite-cell solution computed with a separately written solver. The two implementations share the design formulas and sparse factorization library.

Of 68 guarded static bounds or gap edges, 57 were sound and 11 marginal. Eight marginal values were density bounds, and two of these, both lower bounds, lay above the reference value by less than 0.01\%, inside its uncertainty $u=0.02\%$. Of 24 claims certified under the original rules, 23 agreed and one marginal maximum-density claim was inconclusive against the reference uncertainty. The released feature rule withholds one of the agreeing claims, the gap of the thin resonator with 0.061$a$ ligaments, which span 2.4 voxels at $n=40$. The claims come from six cells and are not independent, so we give no rate bound.

The largest discrepancy among the extrapolated static moduli was 29.6\% for the shear modulus of the 0.061$a$-ligament resonator. The checker marked that quantity unconverged and the cell's certificate did not rely on it. The reference study passed its comparison rule after finer reference levels were added following initial comparisons.

\subsection{Gap checks}

\paragraph{Published ball-and-beam cell.}
For the cell of Lucklum and Vellekoop \citep{lucklum2018bandgap}, we give frequencies in the published normalization $fa/c_l$, where $c_l$ is the longitudinal wave speed of the solid. The paper does not state a Poisson ratio. Assuming an isotropic solid and taking the positive root, the reported modulus, density and $c_l$ give $\nu_s=0.35$, so that $f^{*}=2.08\,fa/c_l$.

\begin{figure}[!tb]
\centering
\includegraphics[width=0.93\linewidth]{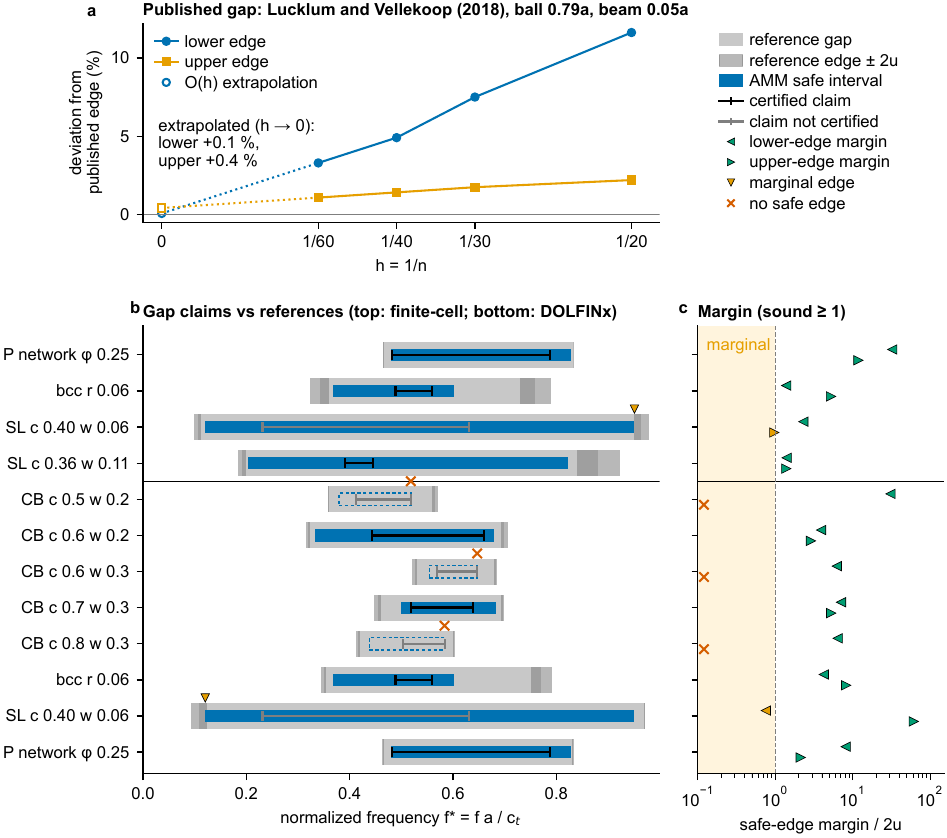}
\caption{Band-gap verification. (a) Mesh-refinement error relative to the published ball-and-beam gap edges, including the extrapolation from the two finest meshes. (b) Claimed intervals, safe intervals of the released guard and reference intervals for the gap claims of the quadratic finite-cell comparison (above the line) and the comparison with the DOLFINx finite-element library (below). Dark gray marks the reference value $R$ plus or minus twice its numerical uncertainty $u$ at each reference edge. Gray claims are not certified under the released rules; crosses mark upper edges without a safe value, whose dashed outline spans the safe lower edge to the claim. Cell labels: CB, cube and bars with cube edge $c$ and bar width $w$; SL, sphere and ligaments; P, Schwarz P network with solid fraction $\phi$; bcc, body-centered-cubic struts with radius $r$. Triangles denote marginal guarded edges. (c) Margin between guarded and reference edges in units of $2u$, positive toward the gap interior. The displayed edges include 19 sound, two marginal and three without a safe value; three cells appear in both comparisons. $f^{*}=fa/c_t$ is the dimensionless frequency of Eq.~\eqref{eq:scale}, and $h=1/n$ is the grid spacing for $n$ voxels per cell edge.}
\label{fig:gap}
\end{figure}

At $n=60$ the computed edges lay 3.3\% (lower) and 1.09\% (upper) above the published values. Extrapolated from $n=40$ and 60 under a first-order assumption, the upper edge was $fa/c_l=0.4722$ versus $0.4702$ published, a difference of $+0.43\%$, and the lower edge rounded to $fa/c_l=0.0437$ in both calculations (\fig{fig:gap}a). Over the orders the guard admits, 0.25 to 2, the same levels give $-11.8$ to $+2.0\%$ (lower) and $-2.0$ to $+0.8\%$ (upper), so the agreement rests on the assumed order.

We used this cell to calibrate the first-order form of the edge guard. Under that form, a deliberately near-edge interval of $fa/c_l=0.055$--0.478 passed the counts at $n=20$ and 30, but its upper endpoint exceeded the guarded edge, and an interior interval of $fa/c_l=0.055$--0.46 passed both checks. The coarse counts alone would have admitted the first interval.

With the current three-resolution guard, both intervals still passed the counts, but the checker left both undetermined. The lower-edge sequence converged slowly (fitted order 0.35), giving a safe lower edge of $fa/c_l=0.0566$, above the claimed 0.055 and the published 0.0437. For the interior interval the upper-edge estimates, 0.4806, 0.4785 and 0.4769 at $n=20$, 30 and 40, neither converged at an admissible order nor stayed within phase noise, and neither interval received a safe upper edge. The 0.05$a$ beams also span only two voxels at $n=40$, below the feature rule. On its calibration cell the current guard therefore abstains instead of accepting either interval.

\paragraph{Independent finite-element comparison.}
We cross-checked the gap calculation on eight cells against a comparison implementation built with DOLFINx and its multi-point constraints extension (DOLFINx-MPC) \citep{baratta2023dolfinx,dokken2026dolfinxmpc}. Five cells use exact geometry and three curved cells share \amm{}'s voxel occupancy. On those three cells the comparison tests the elements, quadrature, Bloch conditions, eigensolvers and convergence with resolution, but not \amm{}'s geometry representation, because both codes use it.

Under the amended procedure and the released guard, 12 of 16 edges were sound and one was marginal. The remaining three upper edges had no safe value (\fig{fig:gap}b,c). The original gap guard had graded as sound on all three. The added reference level and the change to the extrapolation step measure were decided after the first comparison numbers existed. The lower edge of the 0.061$a$-ligament resonator was marginal under all reported variants.

Before the comparison runs, the protocol defined three grades for the study: strong, adequate and weak (\suppref{Section}{S6}{sec:supp-records}). Any unsound edge, failed identity test or claim outside the reference interval made the study grade \emph{weak}. In the identity comparison on a common discrete model, the frequencies agreed to within $1.9\times10^{-9}$ relative against a tolerance of $10^{-6}$. The relative eigenvalue residuals had to be at most $10^{-8}$ on both sides, but the maximum reached $9.67\times10^{-8}$ for DOLFINx and $3.23\times10^{-8}$ for \amm{}. The residuals exceeded the tolerance, so the study grade is \emph{weak}.

The frequency difference in the identity test was nearly five orders of magnitude below the smallest relative reference uncertainty at a compared edge, $1.6\times10^{-4}$. That comparison gives no bound on solver error at each edge. For a symmetric definite pencil, each computed eigenvalue lies within its residual norm, measured in the inverse mass-matrix norm, of an exact eigenvalue \citep{parlett1998symmetric}; the stored residuals use a different norm, so we did not evaluate the bound at the compared edges.

\paragraph{Denser wave-vector sampling.}
A complete gap must hold at every wave vector. The checker counts modes at the 37 sampled wave vectors of Section~\ref{sec:gapcheck}. As a control, we counted Bloch eigenvalues below both claim endpoints on a regular grid over the irreducible wedge $0\le k_z\le k_y\le k_x\le\pi$, with step $\pi/20$ and faces and edges included, giving 1,771 wave vectors per cell. The scan covered the twelve gaps certified under the original gap guard: the eight comparison cells, the resonator with 0.109$a$ ligaments from the finite-cell comparison and the three engineering designs. Counts were taken at the lowest certificate resolution, $n=20$, at grid phase zero.

All 21,252 wave vectors gave the same counts as their certificates. The scan therefore found no mode of the discrete model inside a claimed interval at a grid point. On this grid one band of the 0.109$a$-ligament resonator exceeded its sampled edge estimate by up to 0.30\%, still inside the safe edge, so the sampled set does not always locate the extrema (\suppref{Section}{S6}{sec:supp-records}). The scan tests wave-vector sampling only: it does not bound the discretization error, and points between grid nodes remain unsampled.

\paragraph{Effects of the revised gap rules.}
Recomputing the original edge guard from the stored estimates reproduced the safe edges of all twelve gaps. Fourteen of the 24 edges used the phase-noise branch, in which the resolution changes lie within the phase spread and the spread, up to 4.6\% of the edge frequency, enters the error. Nine of these fourteen sequences drifted monotonically with resolution, and four of the nine, all upper edges, drifted toward the claim, so the original gap guard could treat a slow drift as noise. We therefore changed the rule so that such an edge must pass the order fit, and added the feature rule (Section~\ref{sec:gapcheck}).

With the revised rules, six of the twelve gaps remain certified on the stored solves. Three cube-and-bar comparison cells and the PA12 filter lose their safe upper edge. The 0.061$a$-ligament resonator and the 316L filter fail the feature rule, with 2.4 voxels across the connector at $n=40$. Section~\ref{sec:filter} reports a finer gap check of both filters. On the three cube-and-bar cells the earlier safe upper edge had lain inside the reference gap, so the change withdraws certificates that the reference supports; with the 316L filter and the 0.061$a$-ligament resonator, whose claims also lie inside their references, five of the six withdrawn gaps have a supporting reference. The stricter rules therefore trade yield on these cells for a risk, a drift that carries an edge past the claim, that no cell here showed.

On the 15 distinct edges with a reference and a safe value, 13 from DOLFINx and the two of the 0.109$a$-ligament resonator, which only the finite-cell solver checked, the safe edge lay inside the reference gap by 1.0--40.3\% of the gap width. The released rules changed no verdict on the 50 verification cells. Applied to the stored solves, a Bloch order cap of one instead of two would change no verdict of the twelve gaps, the engineering certificates or the finer filter checks, and one of the 46 benchmark certificates with a gap target (\suppref{Table}{S8}{tab:rules}).

\paragraph{Sensitivity to geometry sampling.}
Each voxel's solid fraction comes from $s^3$ sub-samples with $s=\max(2,\lceil96/n\rceil)$, so the sub-sample grid $ns$ changes irregularly with the resolution, between 96 and 192 per cell edge over the certificate resolutions. On the finer gap ladder of Section~\ref{sec:filter}, $n=30$, 40 and 60, it stays at 120, so the remaining error there omits geometry sampling altogether. The extrapolation therefore mixes mesh refinement with changes in geometry sampling rather than refining both together, and its limit is that of this sampling schedule. Doubling the target sub-sample density at the finest level of the three engineering certificates lowered $\E$ by 0.19--1.06\%, at most 0.18 of the remaining error $e$, and moved each gap edge at $n=40$ by at most 0.42 of its remaining error; the largest of these moves took the 316L lower edge away from the claim.

\paragraph{Benchmark decisions after the rule changes.}
Across the 85 unique certificates of the search benchmark (Section~\ref{sec:search}), the released checker certified 46, refuted one and left 38 undetermined; the original rules had certified 61. Of the 15 withdrawn certificates, ten failed only the feature rule, with load-path features 0.060--0.072 of the cell edge, and five lost a safe upper edge, two of which also failed the feature rule. Of the 38 undetermined certificates, 19 had a gap edge without a safe value, ten an unresolved feature, three a gap claim inside the edge guard band, two modes inside the claimed interval at every resolution, two count failures and two a stiffness interval that straddled its target.

\subsection{Held-out gap test}
\label{sec:heldout}

The comparison cells above also informed the rule changes, so we tested the released gap guard on cells that played no part in them. The protocol was written and hashed before any cell was drawn, and the checker was not changed afterwards. We drew ten cube-and-bar cells, three with thin bars and seven regular ones. Each received one interior and one near-edge claim based on its search-tier gap (\suppref{Section}{S6}{sec:supp-records}). The released checker ran at its defaults. Exact-geometry DOLFINx references were built as in the comparison above, from levels $n=15$, 20 and 30. A claim is true if it lies inside the reference gap narrowed by $2u$ at each edge, false if it extends beyond the reference gap widened by $2u$, and indeterminate otherwise. The protocol set two conditions for passing. The checker had to avoid certifying any false claim. Every safe edge it reported also had to be sound or marginal, even if it left the claim undecided.

\begin{table}[tbp]
\small
\caption{Held-out gap test on ten cube-and-bar cells with two claims each. Safe edges are guarded edge estimates reported for the 20 claims and graded against the references; the other 34 edges had no safe value. The feature rule requires three voxels across the bar at $n=40$.}
\label{tab:heldout}
\tablefit{Xrrrrrr}
\toprule
 & & \multicolumn{3}{c}{Claims by reference truth} & \multicolumn{2}{c}{Safe edges} \\
Cells & Feature rule & False & True & Indeterminate & Sound & Unsound \\
\midrule
Three thin-bar cells & Fails & 5 & 0 & 1 & 1 & 1 \\
Seven regular cells & Passes & 12 & 2 & 0 & 4 & 0 \\
\midrule
Certified & & 0 & 1 & 0 & & \\
\bottomrule
\end{tabularx}
\end{table}

The held-out test failed because one of the reported gap edges was unsound (Table~\ref{tab:heldout}). Of the 20 claims, 17 were false. The checker left all 17 undetermined. For 13 of them, it found modes inside the claimed interval at every resolution.

The edge failure was separate from those acceptance decisions. Of the six graded edges, the failed one was the upper edge for the interior claim on a thin-bar cell with $c=0.619$ and $w=0.066$, whose bars span 2.6 voxels at $n=40$. The edge estimates were 0.741, 0.644 and 0.637 at $n=20$, 30 and 40; the fitted order of 6.0 was capped at two, and the safe upper edge was $f^{*}=0.607$. The reference upper edge is 0.560 with $2u=0.031$. On exact geometry with conforming elements, the finest reference value is itself an upper bound on the band at its wave vector, so the safe edge overstates the gap edge by at least 8\%. The checker left the claim undetermined for two reasons. The lower edge had no safe value, and the bars failed the width check. The reported upper edge of 0.607 still counted toward the test outcome. An order cap of one would have given 0.567, still above the bound, and neither rule alternative of \suppref{Table}{S8}{tab:rules} changes a held-out verdict.

The other five safe edges were sound. Four belong to regular cells, which pass the feature rule, and one is the lower edge of a second thin-bar cell. The checker reported an unsound edge even though its feature rule withheld acceptance. A guarded edge reported for an under-resolved cell should be treated as unresolved. The protocol did not make this distinction, so it is an observation after the result, and four sound edges on regular cells do not establish a rate. We left the released checker unchanged, because a rule changed after this test would need a new held-out set.

Only two claims were true, both interior claims on regular cells, so the test says little about yield. One was certified, with margins of 19.0 and 2.2 times $2u$ and a safe interval that gives up 7.8\% of the reference gap. The other, on a nearly identical cell ($c=0.685$ against 0.688), was left undetermined because its upper edge drifted toward the claim without an admissible order. The claims were mostly false because the search-tier gaps that set them lay above the references, by 13--34\% at the lower edge and 6--146\% at the upper edge, so that even the interior claim was false on seven of the ten cells. The 37 sampled wave vectors missed no extremum that the reference's own wave-vector search found.

\section{Capability and interface acceptance}
\label{sec:capabilities}

The acceptance matrix applied 159 distinct checks in 443 rows across the library, a direct MCP client, the browser application programming interface (API), automated browser clicks and the command line. It covered target types and combinations, family restrictions, physical units, invalid inputs, stale evidence, certificates and exports.

Of the 443 rows, 267 passed, 144 refused as expected and 32 failed their original expectation. Test design or expectations accounted for 31 failures. The remaining failure exposed a defect in a server health report, which we fixed after the run (\suppref{Section}{S3}{sec:supp-interface} and \suppref{Table}{S3}{tab:interface}). All 170 sessions from the acceptance run replayed exactly in the recorded environment, and the PA12 static certificate was identical across the MCP client, browser API, command line and a tutorial. The browser checks used scripted actions without study participants.

\section{Engineering workflows}
\label{sec:cases}

The mount and filter workflows take a specification through design selection, nominal certification, export and reanalysis. Acceptance thresholds were specified before solver runs. Targets were selected using the atlas, and the recorded procedural amendments were applied. Scripts executed the numerical workflows without an agent. A browser walkthrough separately exercised the mount sequence and reproduced the script's certificate. The cases demonstrate the workflow; the evidence that the checker changes acceptance decisions is in Section~\ref{sec:verification}.

\begin{figure}[!htbp]
\centering
\includegraphics[width=0.91\linewidth]{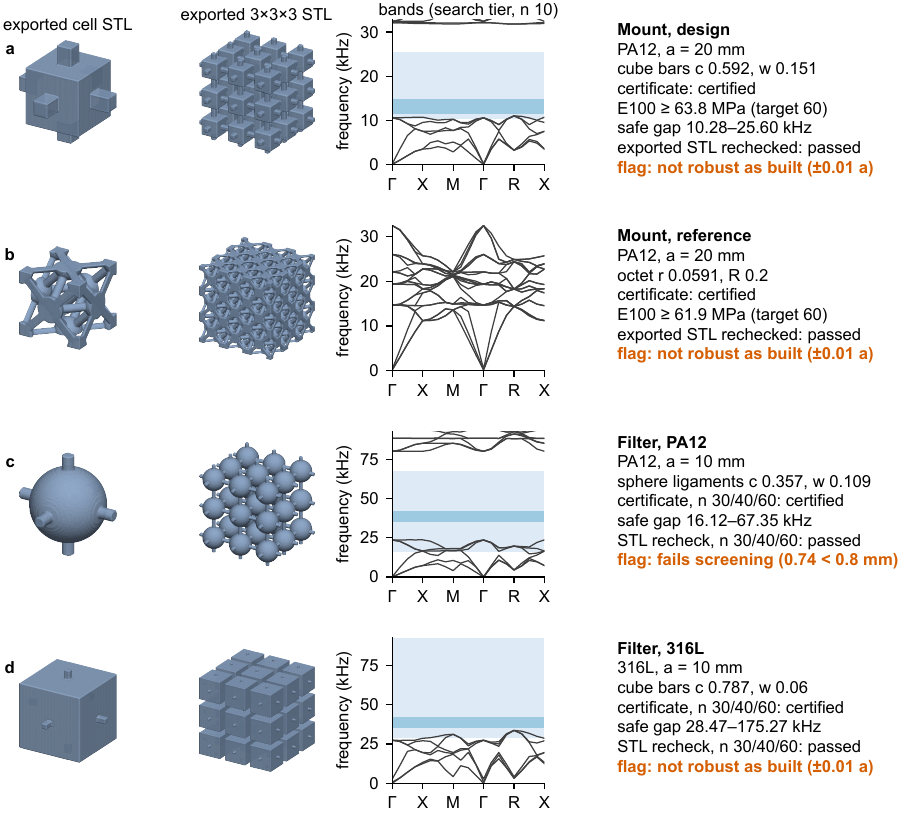}
\caption{Engineering cases and their exported geometry. Each row shows the cell, a $3\times3\times3$ tiled block, the search-tier band diagram and the numerical evidence. Dark and light bands show the requested interval and the guarded safe interval. The plotted bands are search-tier estimates ($n=10$), which overstate frequencies, so some enter the safe interval; the certificate counts no mode inside the requested interval at its sampled wave vectors. The filter certificates are those of the finer gap check (counts at $n=30$ and 40, edges at $n=60$): both filters are certified, and both exports were rechecked at these resolutions. The octet is the stiffness-matched mount reference, selected by screening for a lack of gap in the target band. PA12, polyamide 12; 316L, austenitic stainless steel; STL, stereolithography surface mesh. Geometry parameters, in cell units: $c$ and $w$ are the core size and connector width, and $r$ and $R$ the strut and nodal-sphere radii.}
\label{fig:gallery}
\end{figure}

\subsection{A load-bearing PA12 mount}

The resonator met the mount's nominal targets in simulation. The exported mesh also passed reanalysis, but erosion of the nominal geometry reduced stiffness below the requirement.

\paragraph{Requirements and selection.}
We set an illustrative requirement within the reach of the atlas. The mount must support 1.8~kN with at most 0.25~mm axial deflection in a $6\times6\times6$ block between bonded rigid plates. The structural requirement corresponds to 7.2~kN/mm. With a 120~mm block height $H$ and $120\times120$~mm loaded area $A$, the uniaxial estimate $k=\E A/H$ sets a cell target of $\E\geq60.0$~MPa. The frequency requirement is a complete gap over 11.50--14.95~kHz for structure-borne noise. Gap frequencies scale with $c_t/a$, so 20~mm PA12 cells place complete gaps in the kilohertz range; the case addresses kilohertz structure-borne noise, and isolating a much lower band would need larger cells or heavier resonator masses.

With the nominal PA12 properties ($E_s=1.70$~GPa, $\nu_s=0.39$, $\rho_s=1010$~kg/m$^3$) and $a=20$~mm, Eq.~\eqref{eq:scale} maps the request to $\E/E_s\geq0.03529$ and a dimensionless gap covering 0.2956--0.3843. Because $\E$ scales with $E_s$ and the gap frequencies with $\sqrt{E_s}$, the certificate found below keeps its stiffness target for a static modulus of at least 1.60~GPa (0.941 times nominal) and its gap for a kilohertz storage modulus of at most 2.13~GPa (1.251 times nominal), with the other properties fixed. PA12 is viscoelastic and the case uses nominal datasheet properties, so its static and kilohertz moduli each need checking against this window before use.

An initial atlas scan did not find a feasible band for the higher target of 100~MPa with an upper-to-lower frequency ratio of 1.5. The adopted specification, at 60.0~MPa and a ratio of 1.3, had three atlas candidates that cleared the pessimistic search-tier estimate. This screening preceded the solver runs. The workflow evaluated the top five suggestions and attempted certificates in atlas order, with at most three calls. A cube-and-bar resonator passed on the first call. Its guarded Young's modulus was 63.76~MPa, 6.3\% above the cell target, and the guarded gap was 10.28--25.60~kHz (\fig{fig:gallery}). A later exact-geometry reference from the DOLFINx code of Section~\ref{sec:verification} ($n=15$--30) gave $\E=67.3\pm1.0$~MPa ($\pm2u$) from long-wavelength wave speeds and a gap of 9.73--26.91~kHz, which contains the guarded gap with margins of 3.0 and 2.5 times $2u$.

\paragraph{Block stiffness and slab transmission.}
We selected a reference cell to compare transmission at similar axial stiffness. It had the same material and cell size, and its search-tier band calculation showed no complete gap overlapping the requested interval. The selected octet met the static cell target with 61.85~MPa. It passed on the second certificate call after the first candidate was undetermined. The reference comparison tests the difference in normal-incidence transmission for these two designs with similar axial stiffness.

The homogenized block model used the guarded lower Young's and shear moduli and the less stiff of the solutions at the two guarded Poisson-ratio bounds. The lower plate was fixed, the upper plate was displaced axially and the lateral faces were free. Reaction divided by the imposed displacement gave 7.654~kN/mm for the design and 8.097~kN/mm for the reference. Under 1.8~kN those stiffnesses give deflections of 0.2352 and 0.2223~mm, respectively.

To check the homogenized description against an explicitly resolved lattice, we compared $3\times3\times3$ blocks at the same voxel resolution ($n=32$ per cell). The voxel block was 1.24\% stiffer than the homogenized block for the resonator and 0.67\% stiffer for the octet. The finite comparison checks the role of the plates and free lateral faces at this size for axial loading.

Distributing 1.8~kN across the 36 vertical cell columns gives a hand estimate of 50~N per bar and a nominal axial stress of 5.5~MPa in the 3.0~mm bars. The Euler load of the 8.2~mm clamped bar is about 140 times that force. Stress concentrations, creep and fatigue of PA12 were not checked.

\begin{figure}[!htbp]
\centering
\includegraphics[width=\linewidth]{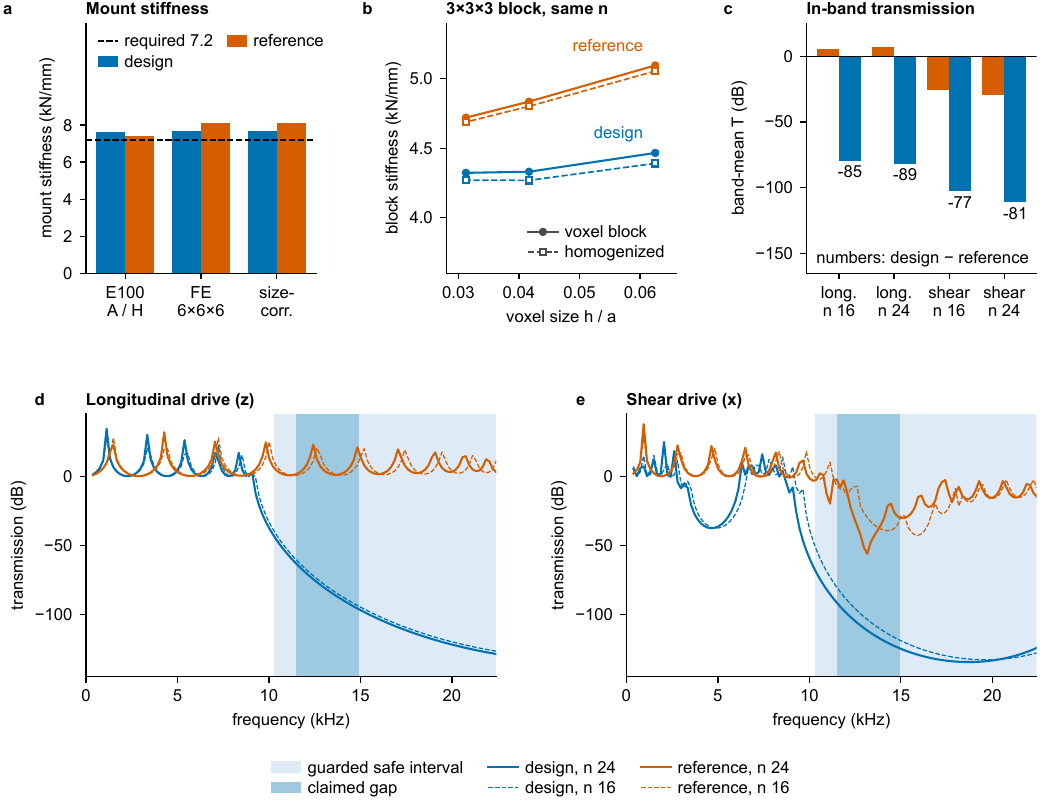}
\caption{Mount checks. (a) Axial stiffness from guarded effective elasticity against the 7.2~kN/mm requirement. (b) The explicitly resolved $3\times3\times3$ block versus the homogenized block at matched voxel resolution. (c) Band-mean transmission of the six-cell slab for longitudinal and shear excitation, with design-minus-reference differences. (d,e) Transmission spectra for normal incidence and loss factor $\eta=0.01$ at two mesh resolutions. Shading marks the claimed and guarded intervals. Values below about $-60$~dB indicate strong attenuation in this model. FE, finite elements.}
\label{fig:mount}
\end{figure}

We also compared transmission through slabs six cells thick, with transverse periodicity, a bonded rigid input plate and a free output face. The prescribed threshold was a band-mean difference of at most $-20$~dB relative to the octet under both longitudinal and shear drive at normal incidence. At $n=16$ the differences were $-84.9$ and $-77.2$~dB, and at $n=24$ they were $-88.9$ and $-81.2$~dB (\fig{fig:mount}). Values below about $-60$~dB are interpreted only as strong attenuation; they are given to show that both meshes stay below that level. The loss factor $\eta=0.01$ is a nominal low value, and a higher PA12 loss would also attenuate the octet, so the margin is specific to this model. The acceptance result is that the slab met the $-20$~dB comparison under the simulated conditions.

\paragraph{Export and surface offsets.}
We converted the exported resonator mesh back to voxels and repeated the checks. It passed both the static and gap targets, with a guarded interval of 10.50--24.87~kHz. The manufacturing screen passed at a 0.8~mm minimum feature limit. Eroding the nominal surface by $0.01a$ reduced $\E$ to 54.4~MPa, below the 60.0~MPa requirement. The reference also failed its erosion check.

\paragraph{Search with a higher stiffness target.}
We raised the nominal stiffness requirement to allow for erosion. We divided the original 60.0~MPa target by the retained fraction $54.4/63.76=0.853$, giving 70.3~MPa, rounded up to 70.4~MPa. The ratio uses the eroded value at $n=32$ and grid phase zero, divided by the guarded nominal value. At the same grid and phase, the nominal cell gave 71.1~MPa and erosion removed 23.6\%. The raised target therefore understates the loss at matched resolution.

The rule set before the run was to check the five best atlas suggestions in order. Release required both a certificate and a passing as-built check under erosion and dilation by $0.01a$. One suggestion was certified, with a guarded $\E$ of at least 74.3~MPa, but its eroded cell reached only 62.0~MPa and its dilated cell failed the gap check. Under the original rules, three were undetermined because the guarded $\E$ interval straddled 70.4~MPa, the gap claim lay inside the edge guard band, or both. The case-1 cell itself was refuted against the raised target, with an $\E$ upper bound of 69.0~MPa.

Every eroded cell fell below 70.4~MPa, at 52.7--65.9~MPa, and the session ended as numerically undetermined without a release. Within this family and these five suggestions, no mount kept both targets through a $0.01a$ surface offset. The as-built values come from one solve at $n=32$, which overstates stiffness relative to the guarded values (71.1 against 63.76~MPa for the case-1 cell), so they screen the suggestions but certify none of them. The released rules change none of these verdicts.

\subsection{A gap filter in two materials}
\label{sec:filter}

The filter specifies a 35.00--42.00~kHz complete gap at $a=10$~mm. We evaluated a PA12 sphere-and-ligament resonator and a 316L stainless-steel cube-and-bar resonator. With the released rules, both needed a finer gap check before the checker accepted them.

\paragraph{Default resolutions.}
Under the original rules both designs and their exports passed (\suppref{Table}{S4}{tab:case}). Applying the released rules to the stored solves withdrew both design certificates. The PA12 upper-edge estimate fell from 72.9 to 72.7 and 70.1~kHz at $n=20$, 30 and 40. The step grew instead of shrinking, so no order fits, and the edge moves toward the claim: it has no safe value although it lies 28~kHz above the claimed 42.00~kHz. The 316L bars span 1.2, 1.8 and 2.4 voxels at the same resolutions, below the three-voxel rule.

\paragraph{Finer resolutions.}
The default gap check stops at these resolutions. In separate runs, we checked both cells with counts at $n=30$ and 40 and edges at $n=60$, in 15 and 39~min on the central processing unit (CPU). The PA12 upper edge fell from 72.7 to 70.1 and 69.3~kHz, a shrinking step, and the certificate holds with a safe gap of 16.12--67.35~kHz. The fitted order of this edge is 4.7, capped at two, so the verdict rests on the 25~kHz between the safe edge and the claim rather than on the order.

On the finer grid the 316L bars span 3.6 voxels, and the upper edge fell from 176.44 to 176.31 and 176.17~kHz. These levels are equally spaced in $h$, so an order of one or less gives steps that do not shrink; the fitted order is 0.93, and the certificate holds with a safe gap of 28.47--175.27~kHz. The order test was corrected after this result exposed it: its first version required a shrinking step, left this edge without a safe value and changed no other verdict when corrected. An exact-geometry reference places the 316L gap at 26.51--176.06~kHz, which contains the safe gap with margins of 3.7 and 19.3 times $2u$.

\paragraph{Export and manufacturing screens.}
Both exported cells passed a separate finer check. Rechecked at these resolutions in 17~min with its earlier wave-vector search, the exported PA12 cell was certified with a safe gap of 15.84--68.46~kHz (upper edge 73.0, 71.1 and 70.3~kHz). Rechecked in the same way in 46~min, the exported 316L cell was certified with a safe gap of 28.58--176.15~kHz. That upper edge lies 0.09~kHz above the reference for the design cell, but the exported mesh is a different geometry whose upper edges lie about 0.8~kHz higher at each resolution, so that reference does not grade it; that edge lies 134~kHz above the claimed 42.00~kHz. The PA12 ligaments span 2.2, 3.3, 4.3 and 6.5 voxels at $n=20$, 30, 40 and 60, which the feature rule accepts.

The PA12 design passed the surface-offset test but failed the screen at its default resolution of 64 voxels per cell, with an estimated minimum solid feature of 0.74~mm against a 0.8~mm limit. Its nominal ligament diameter is 1.09~mm, about seven voxels at that resolution, and the local thickness of the voxel solid, the diameter of the largest inscribed ball, had a minimum of 0.73~mm, so the voxel solid itself is thinner than the design. At 96 and 128 voxels per cell the same screen passed, with estimates of 0.94 and 0.92~mm. The PA12 filter therefore passed the surface-offset test, and the screen only at the finer resolutions.

The 316L design passed screening at its specified 0.3~mm limit, but dilation by $0.01a$ broke the gap counts. Its 0.6~mm bars hold about 0.5\% of the solid, below the 1\% that the screen tolerates, so the reported 1.41~mm feature size describes the cube. The minimum local thickness of 0.47~mm still exceeds the limit.

\subsection{Procedural changes}

Before selecting the mount reference we changed its ranking from raw atlas modulus to modulus corrected by the family's median error. The acceptance rule for the reference was retained. After the first run of the homogenized block model, the finite-block size comparison was changed to use the cell's own tensor at matched resolution, leaving the block stiffness criterion in place. Anti-aliased solid fractions replaced binary ray-parity samples during exported-mesh re-voxelization. This change came after the first static rows of the recheck showed aliasing with the grid phase and before any recheck verdict. The mount sessions, block calculation and mesh recheck were rerun under one code identity after the first case results existed. The original slab calculations were retained. Finally, overlapping GPU processes inflated timings from some runs without changing the recorded numerical results. \suppref{Table}{S4}{tab:case} summarizes the engineering outcomes.

\section{Search benchmark and performance}
\label{sec:search}

On 20 atlas-derived targets, we compared atlas-only selection with Sobol sampling, differential evolution \citep{storn1997differential} and Bayesian search. Each search used 48 new evaluations and one seed. Atlas-only selection used the precomputed atlas without new evaluations. We requested a certificate only for a final candidate that met the targets in the nominal search estimates (\suppref{Section}{S7}{sec:supp-benchmark}).

Under the released rules, Bayesian search certified 12 designs and atlas-only selection ten. An exact McNemar test \citep{mcnemar1947note} gave a Holm-adjusted \citep{holm1979simple} $p=0.50$, so Bayesian optimization showed no clear advantage under the pre-registered rule. With two continuous parameters per family, a dense atlas is close to exhaustive, and the atlas carried most of the certified yield.

Timings were measured on one workstation with an Intel Core i9-13900KF and NVIDIA RTX 4090. Among certified successes, the median time from specification to certificate was 168.0~s (range 95.5--651.8) for full Bayesian search and 106.8~s (54.0--688.2) for atlas-only selection. Across 85 unique benchmark certificates, the median certificate time was 203.9~s (53.6--900.5) with the surface-offset check disabled. These populations differ, so the medians do not decompose into additive workflow components. \suppref{Fig.}{S3}{fig:performance} separates throughput, tool latency, certificate time and solver scaling. \suppref{Fig.}{S4}{fig:atlas} shows the atlas trade-offs and indicates which sampled cells later received a certificate for a specific target.

\section{Optional language-model interface}
\label{sec:agent}

A confirmatory experiment used one run of the OpenAI Codex command-line agent, version 0.165.0 \citep{openai2026codexcli}, with 78 engineering briefs per arm. It compared an assistant given the full tool profile (arm C: bounds, atlas verdict and certificates) with one given specification, search, evaluation and submission tools only (arm B).

On 54 specifications with a decided correct answer, mean per-specification correctness rose from 0.426 to 0.917, a paired difference of 0.491 (95\% confidence interval 0.361--0.620), chiefly through fewer false refusals.

The experiment measures the effect of giving the assistant the tools as a bundle. A specification counts as feasible when a cell passed the frozen checker, which arm C can call and arm B cannot, so the truth labels and arm C's tools share that checker. The frozen checker uses the original gap guard. Under the released rules the candidate cells that established feasibility for three briefs are undetermined. With those briefs excluded, the difference is 0.471 (0.343--0.598) on 51 specifications. \suppref{Section}{S7}{sec:supp-benchmark} gives the design, the amendments and the outcome figure.

\section{Installation, reproducibility and scope}

\paragraph{Installation and replay.}
We installed an earlier build of version 0.2.0 from its wheel in fresh Windows and Linux-container environments. Each CPU test run had 114 passes, three skips, one deselection and zero failures. The browser application runs from a source checkout. Atlas-based queries require the separately versioned atlas and error model. The supported configurations exclude macOS and processors based on the Arm architecture. \suppref{Table}{S5}{tab:identity} identifies the earlier build, the release wheel and the frozen agent-experiment build.

We used the same earlier build for the replay tests. In the recorded environment it reproduced all 15 sampled sessions from the frozen agent experiment. Each session used its original contract version, with the recorded code and checker identities mapped for replay. A replay of the later refusal rule reproduced 17 of 18 calls exactly. The submission decision changed as expected, with zero numeric differences.

A separate comparison tested three sessions in a Linux container. One replayed exactly and two agreed within tolerance. Search and gap calculations were absent from the three sessions, so their cross-environment agreement remains untested. \suppref{Table}{S6}{tab:replay} distinguishes the replay sets and recorded changes.

\paragraph{Scope of the physical model.}
The physics covers small-strain linear elasticity of one isotropic solid in cubic cells. Material presets use nominal datasheet values without scatter. The model excludes anisotropy and porosity from laser sintering PA12 or laser powder-bed fusion of 316L, as well as viscoelastic loss in PA12. A stiffness loss above 5.9\% would remove the mount's nominal margin.

The guarded gaps describe an infinite periodic lattice. They depend on sampled wave vectors and empirical refinement guards. The comparison with published gap data used one cell. A finite block can still carry modes along its surfaces or edges at frequencies inside a bulk gap. The normal-incidence slab test did not check for these modes.

The exact-geometry references cover cube-and-bar cells. They include five cells from the DOLFINx comparison, ten from the held-out test and two engineering cells. The other comparisons share \amm{}'s voxel representation, including those for minimal surfaces, struts and sphere-and-ligament cells such as the curved PA12 filter.

The finite mount checks cover axial static loading and transmission at normal incidence. Strength, buckling, creep and oblique-incidence transmission remain outside those checks. The uniform surface offset and the minimum-feature screen provide separate diagnostics before manufacturing. The validation was numerical, and the interface evaluation used scripted actions.

\paragraph{Scope of the workflow comparisons.}
The search comparison depends on atlas-derived targets and one seed. The assistant experiment used one repeat, so variability between repeats remains unresolved.

\section{Conclusions}

\amm{} separates the search for a periodic cell from the numerical decision to accept it for a specified stiffness or gap. In the near-threshold static test, the checker accepted fewer true claims than an unguarded comparison but avoided the false acceptances seen with that comparison. The gap results show a different limit. In the held-out cube-and-bar test the checker accepted no false gap claim, yet it reported one unsound gap edge for a cell whose bars were too thin to resolve. The test therefore failed its prespecified rule. Reporting an edge and accepting a design are separate decisions.

The engineering cases show why the distinction matters after search. The mount met its nominal targets, but a uniform erosion reduced its stiffness below the requirement. None of the five suggestions for a higher stiffness target kept both requirements under the surface offsets. The two filters needed finer gap checks before their requested intervals could be accepted.

The decision record identifies what a designer needs to revisit. An unresolved estimate can call for a finer calculation. A missed search target can call for exploring a wider family of geometries, and a violated bound can call for revising the requirement. Keeping the nominal decision alongside the surface-offset and manufacturing screens makes it possible to trace a failed check back to the assumption or quantity that needs attention.

\statement{CRediT (Contributor Roles Taxonomy) authorship contribution statement}
Shaoliang Yang: methodology, software, formal analysis, writing -- original draft.\par
\noindent Jun Wang: conceptualization, funding acquisition, supervision, writing -- review and editing.

\statement{Funding}
This work is supported by Santa Clara University.

\statement{Declaration of competing interest}
The authors declare that they have no known competing financial interests or personal relationships that could have appeared to influence the work reported in this paper.

\statement{Data and software availability}
The \amm{} 0.2.0 source, browser application, tutorials, tests, cited result records and scripts for regenerating the tables and data figures are available at \url{\repourl} under the BSD 3-Clause License, as release v0.2.0 with an installable package \citep{yang2026ammsoftware}. The versioned search atlas, its error model and calibration references, and the exported meshes of the engineering cases are attached to the same release under the Creative Commons Attribution 4.0 International License \citep{yang2026ammdata}.

\statement{Declaration of generative AI and AI-assisted technologies in the manuscript preparation process}
During the preparation of this work the authors used Anthropic Claude, at their discretion, to support software development and analysis, and OpenAI Codex to edit and format the manuscript. After using these tools, the authors reviewed and edited the content as needed and take full responsibility for the content of the published article.

\begingroup\small
\setlength{\bibsep}{0pt}\setlength{\parskip}{0pt}\linespread{0.95}\selectfont
\bibliographystyle{unsrtnat}
\bibliography{references}
\endgroup

\FloatBarrier
\bigskip
\appendix
\raggedbottom
\renewcommand{\thesection}{S\arabic{section}}
\renewcommand{\theequation}{S\arabic{equation}}
\renewcommand{\thetable}{S\arabic{table}}
\renewcommand{\thefigure}{S\arabic{figure}}
\setcounter{section}{0}\setcounter{table}{0}\setcounter{figure}{0}\setcounter{equation}{0}
\renewcommand{\theHequation}{supp.\arabic{equation}}
\begin{center}{\Large\bfseries Supplementary material}\end{center}
\addcontentsline{toc}{section}{Supplementary material}
\section{Discretization and acceptance settings}
\label{sec:supp-settings}

Symbols follow the main text: $E_{100}$ and $\nu_{100}$ are the effective Young's modulus and Poisson ratio along $[100]$, $G=C_{44}$ and $K$ the shear and bulk moduli, $a$ the cell edge, $f^{*}=fa/c_t$ the dimensionless form of the physical frequency $f$, $c_t$ the shear-wave speed of the solid and $\nu_s$ its Poisson ratio.

For the implicit solid $\varphi\le t$, with level-set function $\varphi$ and threshold $t$, the code estimates a signed distance $d=(\varphi-t)/|\nabla\varphi|$ near the surface. For a sub-sample of width $\delta$, the coverage is $1/2-d/\delta$ clipped to $[0,1]$. At voxel resolution $n$, the acceptance model uses $s=\max(2,\lceil96/n\rceil)$ samples per coordinate of each voxel.

The periodic mesh uses a consistent mass matrix. Full voxels use nine condensed incompatible modes and $2\times2\times2$ Gauss integration. Cut voxels use midpoint integration of stiffness and mass over the occupied samples. A failed static factorization is retried with a cut-voxel stiffness floor of $10^{-8}$ times the full-voxel stiffness.

For a static quantity, let $m_i$ denote the mean across phases at $h_i=1/n_i$, with $h_1>h_2>h_3$. For a monotone sequence, an order $p$ is fitted to $m_i=m_\infty+Ch_i^p$ over $0.25\le p\le 8$, with limiting mean $m_\infty$ and coefficient $C$, by matching the ratio of successive changes in $h$; with equal steps in $h$, an order of one or less gives changes that do not shrink. The static guard uses $p'=\min(p,1)$. With $r=h_3^{p'}/(h_2^{p'}-h_3^{p'})$, the extrapolated estimate and remaining error are
\begin{equation}
 \widehat m=m_3+(m_3-m_2)r,\qquad e=|m_3-m_2|r.
\end{equation}
If the sequence has no fitted order, the checker treats both changes within twice the finest-grid phase spread $P$ or both relative changes at most 0.001 as noise. It sets $\widehat m=m_3$ and $e=\max(|m_3-m_2|,P)$. Otherwise the quantity is unconverged. The static interval encloses all computed values and $\widehat m\pm2e$. Convergence requires both $e$ and $P$ to be within 0.05 relative to the finest mean, or 0.05 in absolute units for Poisson ratio. The checker uses the guarded lower value for a minimum and the guarded upper value for a maximum. A minimum on $E_{100}$, $G$ or $K$ can be refuted before convergence if the largest computed value misses the target; the main text states the assumption behind this step.

The order cap, the phase set and the factor were examined on 11 cells that are not among the 50 verification cells: flat-faced and thin resonators, thin struts and two minimal-surface cells, solved at $n=32$, 48, 64 and 80 with five grid phases and at $n=96$ with two. For the resonators the converged value lay 2--6\% below the smallest computed value, because the corner and cut-voxel error decays at an order of about 0.5--1 and varies by 2--6\% between phases.

With five phases, order cap one and factor two, no guarded bound lay beyond every limit estimate of a cell, and one or two lay about 1\% beyond the first-order estimate, inside the reference bracket. The median claim margin the guard needed was 4.7\% on the first triplet and 2.4\% on the second. With three phases, six and four bounds lay beyond the estimate, the worst by 3.5\%, and an order cap of two accepted a false claim in the cut-voxel test, so both were rejected. The scoring scripts also read one cell of the 50-cell set, the cube-and-bar cell with cube edge $c=0.7$ and bar width $w=0.15$ (fractions of the cell edge), which had been solved on the same ladder for diagnosis.

For the Bloch edge fit, the order is capped at two. If no order is fitted, the changes are within twice the phase spread and the edge does not move toward the claim, the checker uses the finest estimate and phase spread. A rising lower edge or a falling upper edge without a fitted order has no safe edge, and neither has a sequence that fits neither rule. The finest resolution of either block must also place three voxels across the load-path feature, the diameter of the largest ball that passes through the solid along each axis. It is the bar width for resonators and the strut diameter for strut lattices. For minimal-surface cells it is found on a $128^3$ grid by bisection over the distance to the void, as the largest radius whose interior set still percolates along $x$, $y$ and $z$; on resonators and struts this numeric estimate agrees with the parameter to within about one grid voxel.

For the lower safe edge the checker takes the maximum of the resolution values, the phase values and $\widehat f+2e$. For the upper safe edge it takes the minimum of the same classes of estimates with $\widehat f-2e$.

Candidate extreme wave vectors are located on a coarse grid at $n=10$ with 200 interior samples and the high-symmetry path, followed by six pattern-search rounds in 26 directions from the three best starts per edge. The interior is sampled within $0\le k_z\le k_y\le k_x\le\pi$ in normalized coordinates. At the acceptance resolutions, shift-invert Lanczos iteration \citep{parlett1998symmetric}, through the SciPy \citep{virtanen2020scipy} interface to the ARPACK eigenvalue library \citep{lehoucq1998arpack}, estimates the edge at selected points. An inertia count near the current extreme screens the other points and triggers an edge solve if one can be more extreme.

The static checker accepts the first resolution triplet at which every requested static quantity is converged and its guarded value meets the target. If convergence or margin is insufficient, it tries the next triplet. The final result reports a numerical failure separately from disconnected geometry. A disconnected candidate can be refuted because the load-bearing model requires a percolating solid.

The calibration and demonstration on the published ball-and-beam cell of Lucklum and Vellekoop \citep{lucklum2018bandgap} used an earlier form of the Bloch guard. It took each edge at $n=20$ and 30 over all sampled wave vectors, extrapolated it at first order, $\widehat f=f_2+(f_2-f_1)h_2/(h_1-h_2)$, and moved the extrapolated edge toward the gap interior by $2|f_2-f_1|$. The safe lower edge was the largest of $f_1$, $f_2$ and the moved estimate, and the safe upper edge the smallest. The current three-resolution guard, rerun on the same cell, left both the interior and the near-edge interval undetermined (main text, Section~5.2). For that cell, frequencies are in the published normalization $fa/c_l$, with $c_l$ the longitudinal wave speed of the solid, with $f^{*}=2.08\,fa/c_l$ at the derived $\nu_s=0.35$.

\subsection{Bloch mode count and solver diagnostics}

At each shifted frequency, the count is the negative inertia of $\mathbf K-\sigma\mathbf M$. The mass matrix is positive definite on the retained degrees of freedom. Congruence by $\mathbf M^{-1/2}$ therefore preserves the number of negative eigenvalues of $\mathbf K-\sigma\mathbf M$ and shows that it equals the number of squared eigenfrequencies below $\sigma$. A real block representation doubles the inertia.

The count uses the double-precision symmetric-indefinite $LDL^{\mathsf T}$ factorization of PARDISO, with $L$ unit lower triangular and $D$ block diagonal, and requires even raw counts and no perturbed pivots. The checker also requires exactly three modes below $f^{*}=0.001$ at the zone center.

\section{Documentation comparison and checker settings}

Table~\ref{tab:tools} records the inspected feature evidence for related tools. Partial entries include external solver interfaces, numerical components that require the user to assemble a workflow, or checks that do not furnish per-result acceptance. A documentation comparison can miss features outside the inspected material. For SfePy, the homogenized acoustic-gap example is a building block for band analysis. DOLFINx-MPC (multi-point constraints) provides periodic constraints for a user-specified discretization. The nTop entry refers to its Homogenize Unit Cell block and the documented workflow.

\begin{table}[!htbp]
\small
\caption{Documented features in the inspected sources. Y means documented, P means partial or a component, and N means the feature was not documented in the inspected materials. A question mark denotes an unresolved entry after follow-up inspection. GUI, graphical user interface; TPMS, triply periodic minimal surface; FANS, Fourier-accelerated nodal solvers. Evidence refers to the documentation snapshot used for this study.}
\label{tab:tools}
\setlength{\tabcolsep}{4pt}
\fit{Xcccccc}
\toprule
Tool & GUI & Stiffness & 3D gaps & Checks & Refusal & Evidence \\
\midrule
LatticeWorks & N & P & N & ? & N & \source{https://github.com/mahtab-vafaee/LatticeWorks} \\
MSLattice & ? & N & N & N & N & \source{https://github.com/MSLattice/MSLattice_windows} \\
FLatt Pack & Y & N & N & N & N & \source{https://github.com/ian27ax/FLatt_Pack_dist} \\
TPMS Designer & Y & Y & N & N & N & \source{https://github.com/Alistairj43/TPMS-Designer} \\
TPMSgen & Y & N & N & N & N & \source{https://github.com/albertforesg/TPMSgen} \\
Top6Meta & Y & N & N & N & N & \source{https://github.com/ECM-LAB-NYU/Top6Meta} \\
nTop & Y & Y & N & P & N & \source{https://support.ntop.com/hc/en-us/articles/23019150935955-How-to-run-homogenization} \\
MetamatBench & Y & Y & N & P & N & \source{https://doi.org/10.1145/3711896.3737416} \\
homo3d & N & Y & N & N & N & \source{https://github.com/lavenklau/homo3d} \\
FANS & N & Y & N & P & N & \source{https://github.com/DataAnalyticsEngineering/FANS} \\
fibergen & Y & Y & N & P & N & \source{https://github.com/fospald/fibergen} \\
SfePy & N & Y & P & P & N & \source{https://sfepy.org/doc-devel/examples/phononic-band_gaps.html} \\
DOLFINx-MPC & N & P & P & P & N & \source{https://github.com/jorgensd/dolfinx_mpc} \\
PhoXonic.jl & N & N & P & P & N & \source{https://github.com/hsugawa8651/PhoXonic.jl} \\
AgenticTO & Y & ? & ? & Y & P & \source{https://arxiv.org/abs/2611.04077} \\
This work & Y & Y & Y & Y & Y & Main text \\
\bottomrule
\end{tabularx}
\end{table}

The broader inspection also considered open-source distribution, thermal analysis, agent interfaces and stereolithography (STL) export. Its four other unresolved cells were FLatt Pack and MetamatBench for STL export, nTop for an agent interface and AgenticTO for thermal analysis.

\begin{table}[!htbp]
\small
\caption{Acceptance checks in addition to the guarded target comparisons. A resolution $n$ is the number of voxels per edge. The surface-offset result is a separate flag.}
\label{tab:thresholds}
\fit{p{0.25\linewidth}X}
\toprule
Block & Settings and checks \\
\midrule
Static geometry & Periodic face connectivity in all three directions and no more than 1\% of solid removed as islands. \\
Static solve & All six relative residuals and relative asymmetry at most $10^{-10}$. The smallest normalized tensor eigenvalue must exceed $10^{-8}$. No perturbed pivots. \\
Static phases & $0$, $0.125$, $0.25$, $0.375$ and $0.5$ voxel at each resolution of the static ladder. \\
Gap counts & $n=20,30$. Four points per leg on $\Gamma$--X--M--$\Gamma$--R--X, ten interior samples and the points identified by the coarse search. Equal counts at both endpoints and all sampled wave vectors. Even real-block counts and no perturbed pivots. \\
Gap diagnostics & Three modes below dimensionless frequency 0.001 at $\Gamma$. Solid fractions agree within 0.01 across counted resolutions. The edge-only solve uses $n=40$. Extra phases $0.25$ and $0.5$ are evaluated at the finest counted resolution. \\
Surface offsets & Erode and dilate by $0.01a$. Use the first resolution of the selected static triplet, excluding density targets. Repeat the gap counts at $n=20$ on two points per path leg and four interior points. \\
\bottomrule
\end{tabularx}
\end{table}

\section{Interface checks and engineering cases}
\label{sec:supp-interface}

Table~\ref{tab:interface} preserves the results of the original interface checks, including the rows that prompted corrections. Thirty-one failed rows were due to expectations or test design. An addendum replaced 26 rows with 30 whose expectations were set before rerunning, and all 30 met them. One export predicate was corrected and checked on the existing report. Four family tests expected the wrong atlas row count and were left without a rerun because their decision predicate already held. The remaining failure concerned a health endpoint that reported the mock server identity in live mode; it was fixed after the run. A pre-run probe separately found three library defects, involving a negative shear target, a Zener-ratio target and an incomplete frequency request, and all three were fixed before the matrix ran. Table~\ref{tab:case} gives the geometries and evidence for the engineering cases.

\begin{table}[!htbp]
\small
\caption{Original interface acceptance counts. A refusal that matched the expected response is tabulated separately from a pass. Corrections and reruns are described in the text above.}
\label{tab:interface}
\fit{Xrrrr}
\toprule
Check group & Rows & Pass & Refused & Failed \\
\midrule
Target types and combinations & 63 & 63 & 0 & 0 \\
Geometry families & 36 & 25 & 0 & 11 \\
Materials and units & 85 & 77 & 8 & 0 \\
Invalid requests & 130 & 23 & 97 & 10 \\
Stale certificates and evidence & 32 & 10 & 19 & 3 \\
Certificates, surface offsets and evaluation & 23 & 19 & 4 & 0 \\
Export and screening & 12 & 11 & 0 & 1 \\
Outcomes and tool profiles & 46 & 25 & 15 & 6 \\
Tool lists and server state & 5 & 4 & 0 & 1 \\
Browser clicks & 6 & 5 & 1 & 0 \\
Replay groups & 5 & 5 & 0 & 0 \\
\midrule
Total & 443 & 267 & 144 & 32 \\
\bottomrule
\end{tabularx}
\end{table}

\begin{table}[!htbp]
\footnotesize
\caption{Nominal, exported-geometry and screening results; PA12, polyamide 12. Length parameters are fractions of the cell edge. For cube-and-bar designs $c$ is the cube edge and $w$ the bar width. For the sphere-and-ligament design they are the core radius and ligament diameter. For the octet, $r$ and $R$ are the strut and nodal-sphere radii. A dash indicates that the quantity was not a target of that case. Safe-gap endpoints are guarded edge estimates; only the claimed interval is count-verified. Table~\ref{tab:rules} defines the original and the released rules. The finer gap check and export recheck apply the released rules with counts at $n=30$ and 40 and edges at $n=60$. The minimum solid feature is the smallest opening that removes more than 1\% of the solid, so it does not resolve the 316L bars (0.6~mm wide, about 0.5\% of the solid). Screening used the default 64 voxels per cell; at 96 and 128 voxels per cell the PA12 filter's minimum solid feature was 0.94 and 0.92~mm and its screen passed.}
\label{tab:case}
\fit{p{0.29\linewidth}*{4}{>{\raggedright\arraybackslash}X}}
\toprule
 & Mount & Reference & PA12 filter & 316L filter \\
\midrule
Cell edge (mm) & 20 & 20 & 10 & 10 \\
Design & Cube and bars & Octet & Sphere and ligaments & Cube and bars \\
Parameters & 0.592174, 0.151304 & 0.05913, 0.2 & 0.356522, 0.108696 & 0.786957, 0.06 \\
Guarded $E_{100}$ (MPa) & 63.76 & 61.85 & -- & -- \\
Solid fraction upper bound & 0.236 & 0.218 & 0.198 & 0.490 \\
Safe gap, original rules (kHz) & 10.28--25.60 & -- & 16.51--66.14 & 30.49--175.79 \\
Export recheck gap, original rules (kHz) & 10.50--24.87 & -- & 16.12--68.27 & 29.66--176.73 \\
Certificate and recheck, released rules & Certified & Certified & Undetermined (no safe upper edge) & Undetermined (bars 2.4 voxels at $n=40$) \\
Finer gap check (kHz) & -- & -- & Certified, 16.12--67.35 & Certified, 28.47--175.27 \\
Finer export recheck (kHz) & -- & -- & Certified, 15.84--68.46 & Certified, 28.58--176.15 \\
Exact-geometry reference gap (kHz) & 9.73--26.91 & -- & Not computed (curved) & 26.51--176.06 \\
Surface-offset check & Fail & Fail & Pass & Fail \\
Minimum solid feature (mm) & 2.19 & 2.03 & 0.74 & 1.41 \\
Screening limit (mm) & 0.8 & 0.8 & 0.8 & 0.3 \\
Manufacturing screen & Pass & Pass & Fail & Pass \\
\bottomrule
\end{tabularx}
\end{table}

\subsection{Mount walkthrough}

\begin{figure}[!htbp]
\centering
\includegraphics[width=0.94\linewidth]{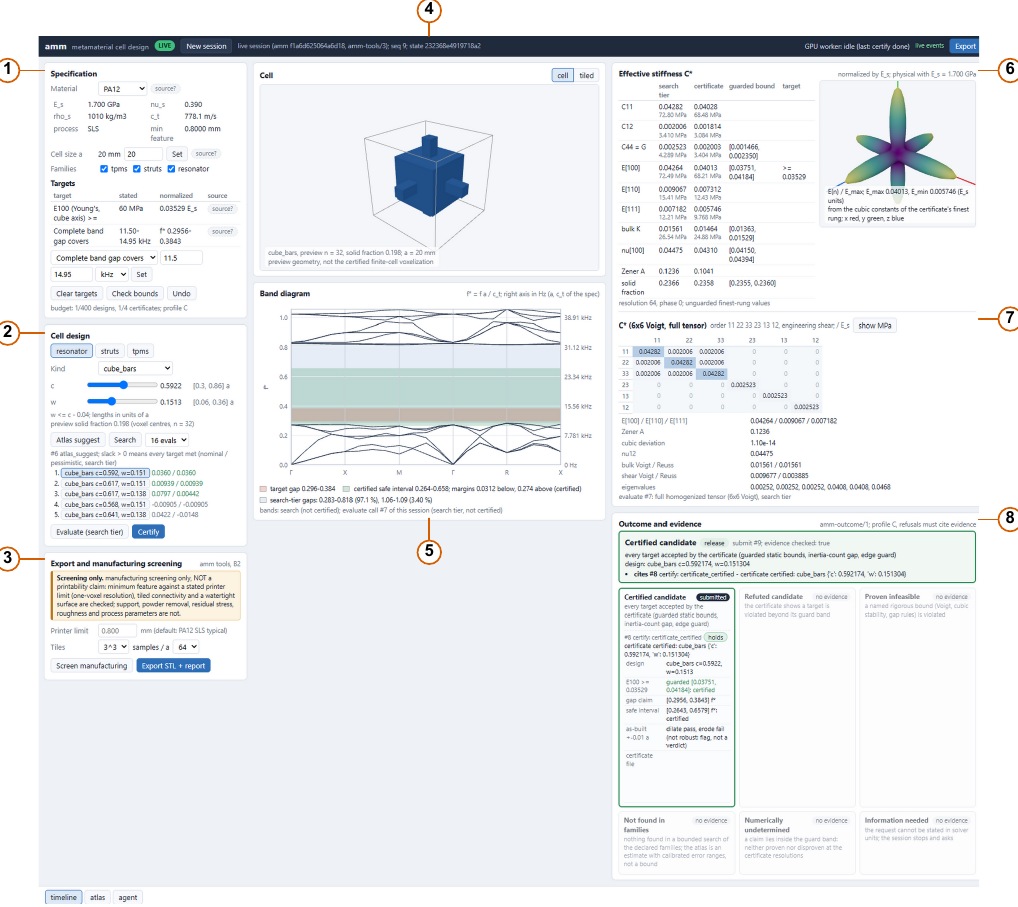}
\caption{Browser application for the polyamide 12 (PA12) mount after certification. Region 1 gives the specification and normalized targets. Region 2 gives candidate design controls and atlas suggestions. Region 3 shows export and screening. Region 4 previews the unit cell. Region 5 shows search-tier bands, the requested gap and the guarded interval. Regions 6 and 7 show effective moduli and the full tensor. Region 8 records the accepted outcome, with guarded $E_{100}\ge63.76$~MPa; its certificate file path is masked. SLS, selective laser sintering.}
\label{fig:gui}
\end{figure}

The application displays the physical specification beside its normalized values and labels search estimates separately from the certificate. Figure~\ref{fig:gui} gives the mount example, with the display regions keyed in the caption. The accepted design and the screen appear in separate regions because they answer different questions about the same geometry.

\subsection{Workflow states}

\begin{figure}[!tb]
\centering
\includegraphics[width=0.80\linewidth]{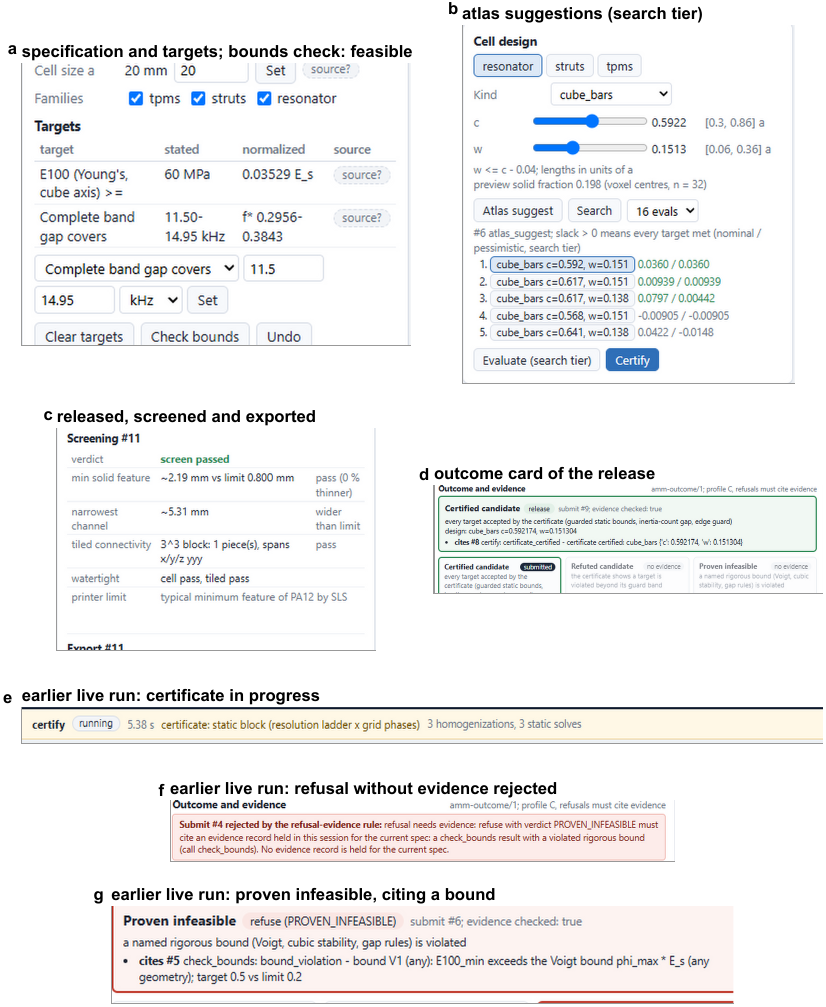}
\caption{Browser actions, cropped from the application window. (a--d) Mount session: specification and bounds check, atlas suggestions, screening and export, and the outcome card of the release. (e--g) In an earlier live session using tool contract version 2, certification is in progress, a refusal without supporting evidence is rejected, and a refusal citing a violated bound is accepted. In (c), the source note of the printer limit is masked. SLS, selective laser sintering.}
\label{fig:steps}
\end{figure}

Figure~\ref{fig:steps} gives the main browser steps for the mount and shows two refusal responses from contract version 2. An atlas assessment can guide the next action. The outcome card appears after submission and points to the record that supports the decision.

\section{Search and timing details}

The primary search targets were drawn without replacement from 42 witnessed confirmatory specifications. Sampling was stratified by witness family and target type (static, gap or combined). Atlas construction, error fitting and search tuning predated the confirmatory specifications. The target construction still favored access to the atlas because every witness came from an atlas row. Four additional calibration targets at $\nu_s=0.342$ and $0.39$ were evaluated outside the hypothesis tests. Under the original rules, the full Bayesian search and the variant without error correction each certified two of the four, with different certified sets; under the released rules each certifies one, the same target.

Exact two-sided McNemar tests \citep{mcnemar1947note} compare final certification and paired sign-flip tests \citep{good2005permutation} compare the mean penalized candidate index. Holm adjustment \citep{holm1979simple} applies over the three baseline comparisons and separately over the two ablations. Identical candidate-and-target pairs share a certificate. The 88 primary certificate requests correspond to 78 unique certificates. Seven unique calibration certificates bring the total to 85.

\begin{figure}[!htbp]
\centering
\includegraphics[width=0.88\linewidth]{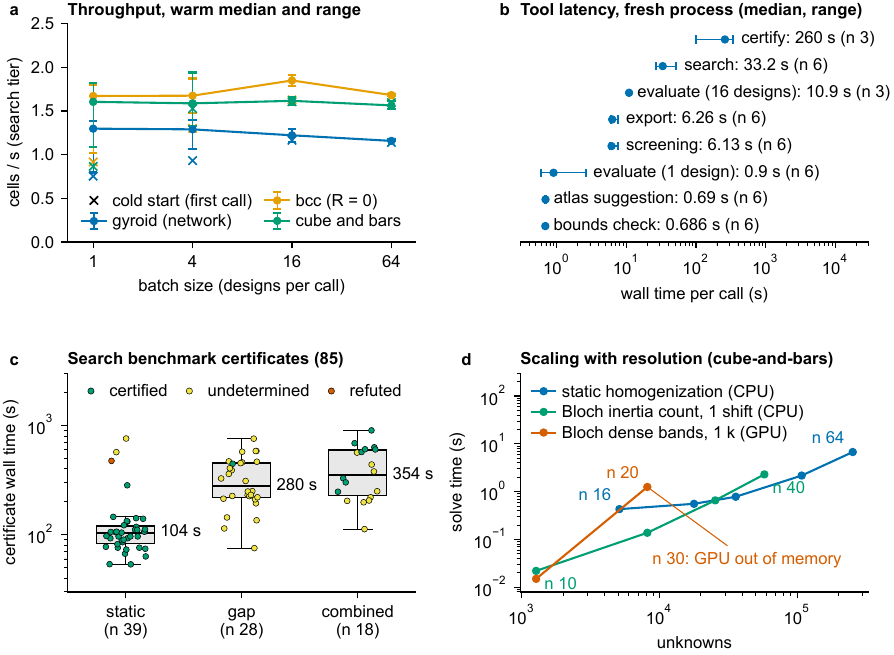}
\caption{Performance on one workstation with an NVIDIA RTX 4090 and Intel Core i9-13900KF. (a) Warm throughput by design family and batch size, with cold first calls marked by crosses. (b) Tool-call latency in fresh processes. Medians below 0.01~s are omitted. (c) Wall time of the 85 unique search benchmark certificates, with the surface-offset check off, colored by the verdict of the released rules. (d) Solve time versus unknowns for static homogenization, one CPU inertia count and dense GPU bands at one wave vector. The dense GPU calculation ran out of memory at $n=30$. $R$, nodal-sphere radius of a strut lattice; bcc, body-centered cubic; CPU, central processing unit; GPU, graphics processing unit.}
\label{fig:performance}
\end{figure}

\begin{figure}[!htbp]
\centering
\includegraphics[width=\linewidth]{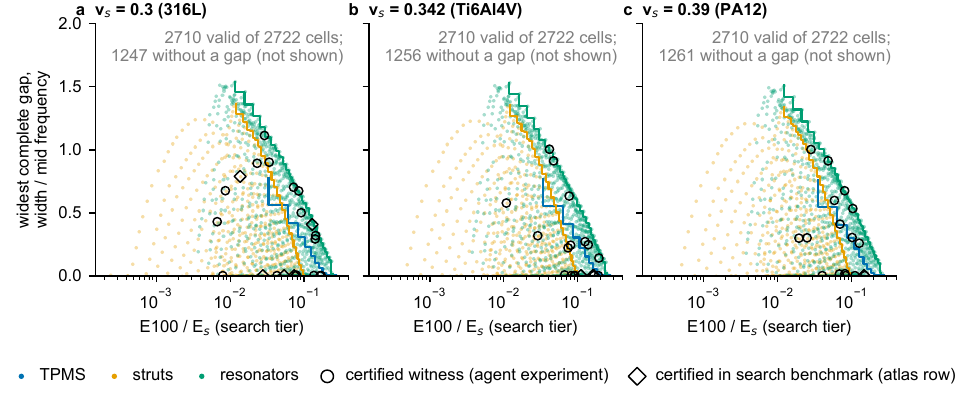}
\caption{Search-tier atlas estimates at three solid Poisson ratios, those of the 316L stainless steel, Ti6Al4V titanium alloy and polyamide 12 (PA12) presets. Each panel plots $E_{100}/E_s$ against the width of the widest complete gap divided by its mid-frequency. Curves join the nondominated cells within each family. Cells without estimated gaps are omitted except where a later certificate is marked at the axis floor. TPMS, triply periodic minimal surface. Circles mark atlas cells certified as witnesses for the specifications of the agent experiment, and diamonds mark other atlas cells certified in the search benchmark. Marker positions use search-tier coordinates. A certificate applies to the requirement against which the cell was checked. $E_{100}/E_s$ is the effective Young's modulus along $[100]$ relative to that of the solid.}
\label{fig:atlas}
\end{figure}

\section{Agent experiment and replay}

The confirmatory run used the OpenAI Codex command-line agent \citep{openai2026codexcli} with GPT-6.1-Sol and high reasoning effort. The 78 briefs per arm include paired forms of some specifications. The analysis clusters by underlying specification. It compares 54 specifications with decided truth and separately reports unresolved specifications and clarification. A witnessed feasible specification has a cell that passed the frozen checker. A proven refusal is evaluated against a bound. Unresolved truth allows abstention. The primary test uses a paired sign-flip permutation \citep{good2005permutation} and the interval is a cluster bootstrap over specifications \citep{field2007bootstrapping}.

At the brief level, the model-only arm released zero designs, the search-tools arm released 15 and the full arm released 52. The scorer checked submitted designs against frozen numerical criteria, including for the arms without a certificate tool. The corresponding counts of false refusals were 55, 43 and six. At specification level, the mean false-refusal rate among 42 witnessed specifications fell from 0.738 to 0.107 when moving from B to C, a difference of 0.631 with a 95\% confidence interval of 0.500--0.762. These brief and specification counts use different units.

\begin{table}[!htbp]
\small
\caption{Frozen identities and stopping record for the confirmatory experiment, and the identities of the released version. The two amendments did not change the four frozen identities.}
\label{tab:identity}
\fit{p{0.26\linewidth}X}
\toprule
Component & Recorded identity or action \\
\midrule
Tool code & \texttt{90f3823f15cbeff4} \\
Checker & \texttt{f2e482d81382b400} \\
Brief set & \texttt{194701c05dce17d2} \\
Atlas and error model & \texttt{de79a79e98e88731} \\
Amendment A1 & After 29 of 234 jobs had finished, dropped two confirmatory repeats, development runs and the second provider. Updated the harness and protocol. \\
Amendment A2 & After the retained jobs had finished and before reading or scoring any confirmatory outcome, dropped the oracle-specification arm. \\
Scored phase & 234 jobs in one repeat. Of these, 205 used the amended harness: 68 each in arms B and C and 69 in arm D. \\
Released tool code & \texttt{bb1a90cdc52b7f93}; the earlier build used for the installation and replay tests has tool code \texttt{f1a6d625064a6d18} and checker \texttt{c5f8c89871b3bad6} \\
Released checker & \texttt{f99ce2b2f65d7630}; original rules \texttt{f141e46cb67a78db} and \texttt{c5f8c89871b3bad6} (Table~\ref{tab:rules}); drift and feature rules before the order-test correction \texttt{a900aa3f152174bc} \\
\bottomrule
\end{tabularx}
\end{table}

\begin{table}[!htbp]
\footnotesize
\caption{Replay sets. Exact comparison excludes timing. One Linux-container session was classified as within tolerance only after a diagnostic rerun traced the remaining difference, the hash of an export report, to floating-point round-off. Sessions written under an older contract version are replayed through a translation to that version. The Linux-container comparison tests outcomes and numeric agreement within tolerance.}
\label{tab:replay}
\fit{Xrrrr}
\toprule
Replay set & Sessions & Exact & Within tol. & Changed \\
\midrule
Frozen agent sample & 15 & 15 & 0 & 0 \\
Targeted refusal replay & 1 & 0 & 0 & 1 \\
Engineering cases and walkthrough & 7 & 7 & 0 & 0 \\
Earlier tutorials & 2 & 2 & 0 & 0 \\
Earlier browser sessions & 3 & 3 & 0 & 0 \\
Earlier smoke tests & 5 & 5 & 0 & 0 \\
Linux container & 3 & 1 & 2 & 0 \\
\bottomrule
\end{tabularx}
\end{table}

\section{Data records and borderline cases}
\label{sec:supp-records}

The data manifest identifies the versioned atlas and error-model bundle. The atlas stores one line per geometry and Poisson ratio in newline-delimited JavaScript Object Notation (JSON). Each row records family, parameters, validity, normalized static quantities and sampled gaps. The error model gives family-specific ranges, thin-feature rules and the calibration domain. Table~\ref{tab:data} maps the result records to the quantities they retain.

\begin{table}[!htbp]
\footnotesize
\caption{Data records for reproducing the reported comparisons. Normalized quantities use the scales in the main text.}
\label{tab:data}
\fit{p{0.23\linewidth}X}
\toprule
Record & Contents \\
\midrule
Registry parity & Rebuilt and stored solid fraction, tensor norm and relative difference on the common mask. \\
Reference comparisons & Geometry, reference levels, estimate and uncertainty, guarded value, edge or bound classification. \\
Interface matrix & Check group, interface, expected and observed response, and original row verdict. \\
Search benchmark & Specification and method, final candidate and index, nominal slack, certificate result and elapsed time. \\
Engineering cases & Material and geometry, target, certificate, block stiffness and deflection, slab transmission, exported geometry and reanalysis. \\
Agent experiment & Frozen brief and checker identities, outcome scores per brief and per-specification effects under amendment A2. \\
\bottomrule
\end{tabularx}
\end{table}

For the higher-order reference comparisons, the estimate $R$ is based on three reference levels. When the sequence is monotone with shrinking steps, the reference extrapolates from the finest level with the fitted order clipped to $[0.5,4]$. The uncertainty $u$ is the extrapolated remaining error with a floor of $2\times10^{-4}$ times the magnitude of the finest value in the quadratic finite-cell comparison, or $10^{-4}$ times the finest value in the DOLFINx band-edge comparison. For irregular sequences the finest value supplies $R$ and the largest change between adjacent reference levels supplies $u$. The resulting uncertainty band reflects numerical resolution in the reference calculation.

In the six-cell comparison, the marginal maximum-density claim for a Schwarz P sheet cleared the internal bound only narrowly, while the external reference uncertainty made the comparison inconclusive. The marginal lower edge of the 0.061$a$-ligament resonator changed sides when the quadratic reference levels were refined. In the DOLFINx comparison its lower edge stayed marginal under all variants. Under the originally planned reference levels, the six-cell comparison gave 48 sound and 20 marginal values, and the DOLFINx comparison with the original gap guard gave four marginal edges; neither had an unsound value.

The pre-registered rule of the DOLFINx comparison had three grades. \emph{Strong} required every identity test to pass, no unsound edge, at most one marginal edge, every claim inside the reference interval and no reference band inside the safe interval at any wave vector. \emph{Adequate} allowed more than one marginal edge, or $2u$ larger than the margin on three or more edges. Any unsound edge, failed identity test or claim outside the reference interval made it \emph{weak}.

We also compared the band extrema on the dense scan grid of the twelve gaps (main text, Section~5.2) with the edge estimates from the 37 sampled wave vectors at $n=20$. In 11 gaps no grid point went beyond them. In the 0.109$a$-ligament resonator the lower band exceeded its estimate by at most 0.30\% at 12 grid points off the high-symmetry path, still below the safe lower edge and the claim. The search benchmark also retained final candidates that missed the nominal threshold and candidates that reached the checker but were undetermined.

\subsection{Supplementary verification tests}

In the main-text near-threshold tests, the checker certified 21 true claims that the first-order band rejected, 19 of them in the phase-noise branch, where the checker's band is narrower, and that band accepted six true claims that the checker rejected, all with fitted order below one, where the checker's band is wider. The three false acceptances under the band with fitted order and factor 1.25 were shear moduli with observed orders 2.9 and 4.4, which narrow the band, whereas the checker caps the static order at one.

A static rule like the gap guard's, under which a quantity whose phase means drift toward the target on the phase-noise branch counts as unconverged, would have withdrawn 48 of the 454 true certifications and no false one, and ten of the 85 benchmark certificates; the released checker does not apply it (Table~\ref{tab:rules}).

Five cells had stored gap edges at three resolutions. The checker certified none of the 50 false gap claims and 3 of the 23 true claims. The unguarded rule accepted two false claims. Both grid convergence index bands accepted no false claims, and they accepted 13 and 1 true claims, respectively. The stored bounds come from the rung each certificate used, so a finer rung could certify more true claims.

\paragraph{Sampling for the held-out test.}
We drew ten cube-and-bar cells with a fixed seed, three with thin bars, of width $w$ between 0.06 and 0.10 of the cell edge, and seven regular ones with $w$ between 0.10 and 0.25, all with a cube edge $c$ between 0.40 and 0.75. A draw was rejected if it lay within 0.02 in both $c$ and $w$ of a cube-and-bar cell used earlier in this study, or if its search-tier gap had a relative width below 0.10.

Each cell received two claims set from its search-tier gap of width $W$: an interior claim that starts $0.25W$ inside each edge and a near-edge claim that starts $0.05W$ inside. We ran the released checker at its default settings.

Table~\ref{tab:rules} lists the checker rule versions and the changes to the comparison procedures in the order they were made. Each later rule version re-judged the stored solves.

\begin{table}[!htbp]
\footnotesize
\caption{Checker rule versions and changes to the comparison procedures, in order. Identities are those of the checker (Table~\ref{tab:identity}).}
\label{tab:rules}
\fit{>{\raggedright\arraybackslash}p{0.25\linewidth}X}
\toprule
Version or change & Trigger and effect \\
\midrule
Calibration form of the gap guard & Edges at the two counted resolutions ($n=20$, 30), first-order extrapolation and factor two, calibrated on the published ball-and-beam cell (Section~S1). \\
Original rules (three builds) & Three-resolution edge guard with the phase-noise branch, before the drift and feature rules. \texttt{f2e482d81382b400}: DOLFINx and finite-element comparisons and the frozen agent experiment; \texttt{f141e46cb67a78db}: search benchmark and six-cell comparison; \texttt{c5f8c89871b3bad6}: engineering cases. \\
Drift and feature rules (\texttt{a900aa3f152174bc}) & Added after recomputing the guard on the twelve dense-scan gaps showed four upper edges drifting toward the claim within the phase noise. Such an edge must pass the order fit, and the finest resolution must put three voxels across the load-path feature. Withdrew six of the twelve gap certificates and 15 of 61 benchmark certificates; no verdict changed on the 50 verification cells or the mount. \\
Order test judged in $h$ (\texttt{f99ce2b2f65d7630}, released) & Corrected after the finer 316L gap check showed that the first test required a shrinking step, which equal steps in $h$ preclude at order one or less. Changed one verdict, that finer check, from undetermined to certified. \\
50-cell classification & The final three-outcome criterion was defined after results existed (main text, Section~5.1). \\
Finite-cell reference levels & Finer static and Bloch reference levels added after initial comparisons; the levels in place before the post-comparison refinements gave 48 sound and 20 marginal values and no unsound one. \\
DOLFINx comparison & A reference level and the measure of the extrapolation step were set after the first comparison numbers; with the planned levels and the original gap guard, four edges were marginal and none unsound. \\
Not adopted: gap order cap one & Applied to the stored solves, one of 46 benchmark certificates with a gap target and six of 66 agent-experiment witness certificates with a gap target change to undetermined; no comparison, engineering, finer-filter or held-out verdict changes. \\
Not adopted: static drift rule & A static quantity whose phase means drift toward a target on the phase-noise branch counts as unconverged. Applied to the stored solves, 48 of 454 true and none of 750 false near-threshold claims lose certification, and ten of 85 benchmark and 12 of 94 witness certificates change to undetermined; no engineering or held-out verdict changes. \\
Held-out gap test (released checker, unchanged) & Protocol fixed before any cell was drawn (main text, Section~5.3). Ten new cube-and-bar cells, 20 claims: no false claim certified; one of six graded safe edges unsound, on a thin-bar cell that fails the feature rule. No rule was changed after the test. \\
\bottomrule
\end{tabularx}
\end{table}

\section{Search benchmark and agent experiment}
\label{sec:supp-benchmark}

\subsection{Search benchmark}

The benchmark compared six methods on 20 specifications with witnesses under the original rules. A witness is an atlas cell that passed the checker for that specification. The specifications were not used in development, but they were derived from atlas cells, which favors atlas-based methods.

The four main methods were atlas-only selection, Sobol sampling, differential evolution \citep{storn1997differential} and the full Bayesian search. Two further variants removed atlas seeding or error correction from the Bayesian search.

\begin{figure}[!htbp]
\centering
\includegraphics[width=\linewidth]{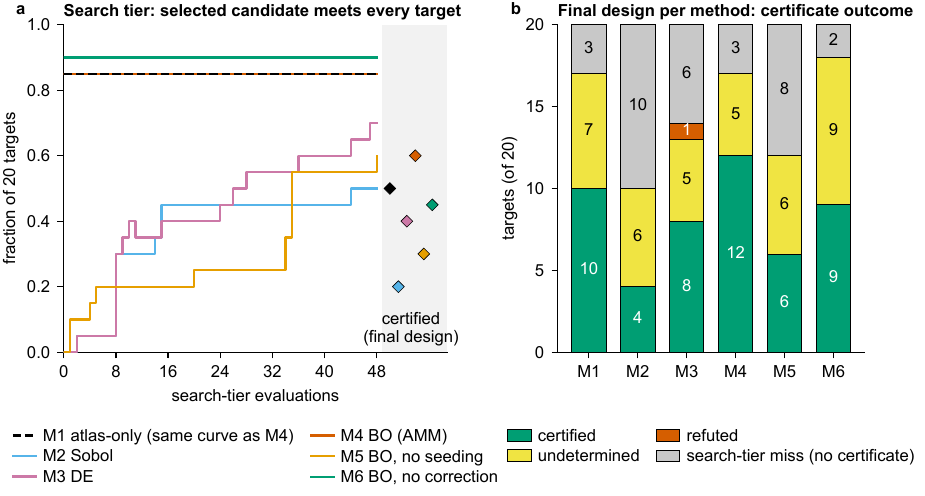}
\caption{Search on 20 atlas-derived targets. (a) Fraction of targets with positive nominal slack for the method's selected candidate as evaluations accrue. Diamonds give the fraction certified after final-candidate checking under the released rules. Intermediate candidates were not certified. The atlas-only and full Bayesian search attainment curves coincide. (b) Final outcome per target and method. A search-tier miss received no certificate. BO denotes Bayesian optimization and DE denotes differential evolution. M1 to M6 number the methods, and AMM (AgenticMetamaterials) is the tool of the main text.}
\label{fig:search}
\end{figure}

All searching methods used the same evaluator and 48 new search-tier evaluations with one seed per method. Atlas-only selection used zero new evaluations. A certificate was requested only for the final candidate with positive nominal slack. As a result, the curves in \fig{fig:search} describe estimated attainment and only the final endpoints give certified yield. All verdicts below are those of the released rules, re-judged from the stored solves. Under these rules the witnesses of two targets are no longer certified, and for one of the two no method produced a certified design; we keep all 20 targets as drawn.

\begin{table}[!htbp]
\small
\caption{Final outcomes for each method on 20 targets under the released checker rules. Undet., numerically undetermined. Cost $C$ is the evaluation index of the final candidate for a certified success, or 48 for a failure. The atlas-only method has index zero on success.}
\label{tab:search}
\tablefit{Xrrrrr}
\toprule
Method & Certified & Undet. & Refuted & Search-tier miss & Mean $C$ \\
\midrule
Atlas only & 10 & 7 & 0 & 3 & 24.00 \\
Sobol sampling & 4 & 6 & 0 & 10 & 44.40 \\
Differential evolution & 8 & 5 & 1 & 6 & 39.05 \\
Full Bayesian search & 12 & 5 & 0 & 3 & 30.15 \\
Bayesian, without atlas seeding & 6 & 6 & 0 & 8 & 42.25 \\
Bayesian, without error correction & 9 & 9 & 0 & 2 & 31.70 \\
\bottomrule
\end{tabularx}
\end{table}

Bayesian search certified 12 designs, atlas-only selection ten, Sobol sampling four and differential evolution eight (Table~\ref{tab:search}); under the original rules the counts were 14, 12, six and nine.

The pre-registered clear-advantage criterion required either a significantly greater certified yield than the best baseline or at least equal yield and a statistically supported reduction of at least half in the cost measure. The paired success comparison against atlas-only, an exact McNemar test \citep{mcnemar1947note}, gave Holm-adjusted \citep{holm1979simple} $p=0.50$. Bayesian search showed no clear advantage under the pre-registered rule. Against Sobol sampling its yield was higher, with all eight discordant targets in its favor (Holm-adjusted $p=0.023$), but Sobol sampling is not the best baseline.

For the cost analysis, a success is charged the index at which its final candidate appeared and a failure is charged the entire budget. This is a retrospective index and the searching methods still spent the full budget. The index cost was lower for Bayesian search than for Sobol sampling (paired sign-flip test \citep{good2005permutation}, Holm-adjusted $p=0.0146$). Its comparison with differential evolution was nonsignificant after adjustment ($p=0.178$). With one seed, an atlas-derived target set and only 20 primary targets, these results leave uncertainty about differences on independently sourced targets. With two continuous parameters per family, a dense atlas is close to exhaustive, so these targets leave search little to add; the atlas carried most of the certified yield.

\subsection{Agent experiment}

We tested whether an assistant given an engineering brief made better decisions when it could call bounds and certificate tools. The completed confirmatory experiment comprises one repeat of the OpenAI Codex command-line agent, version 0.165.0 \citep{openai2026codexcli}, with 78 briefs per arm. Arm D used the language model alone. Arm B used profile~B, with specification, search, evaluation and submission tools. Arm C used the full profile~C, which adds the tools that return analytical bounds, an atlas verdict and certificates.

The primary comparison pairs arms C and B on the same specification and treats the added tools as a bundle. The scorer evaluated releases against the reference targets and material, independent of the specification entered by the assistant.

\begin{figure}[!htb]
\centering
\includegraphics[width=0.76\linewidth]{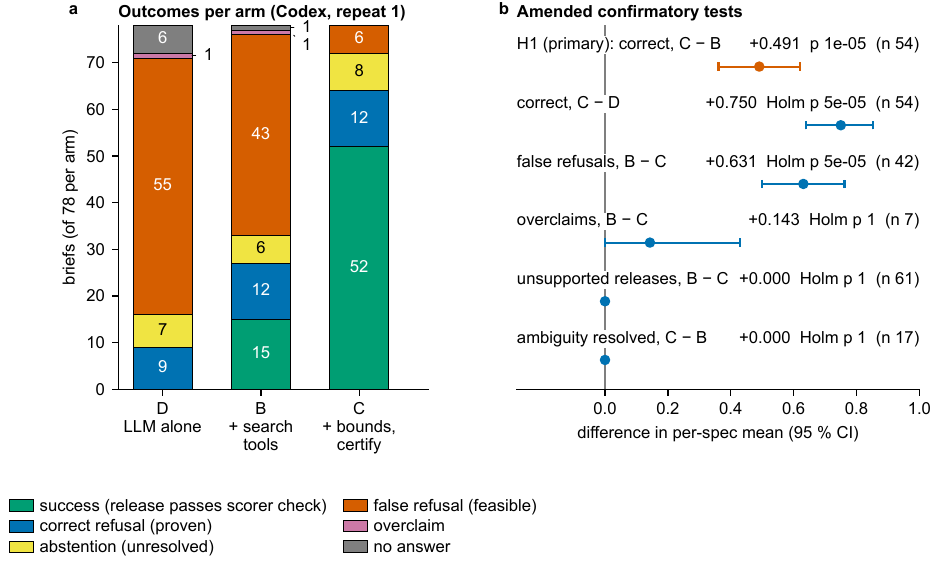}
\caption{Confirmatory experiment with one Codex repeat and the frozen implementation, using the amended analysis recorded in Table~\ref{tab:identity}. (a) Outcomes across 78 briefs per arm; a success is a release that passes the scorer's check, applied after the session in every arm. D is the language model alone, B adds the design tools and C adds bounds and certificates. (b) Paired differences in per-specification means with 95\% cluster-bootstrap \citep{field2007bootstrapping} confidence intervals. Secondary tests are Holm adjusted \citep{holm1979simple}. Brief counts and specification-level effects have different denominators. LLM, large language model.}
\label{fig:agent}
\end{figure}

On 54 specifications with a decided correct answer, the bundle increased mean per-specification correctness from 0.426 to 0.917, a paired difference of 0.491 (95\% confidence interval 0.361--0.620). The gain chiefly reflected fewer false refusals of requests with certified witnesses. Unsupported releases were zero in every arm (\fig{fig:agent}). A zero count here describes the observed sessions, including the model-only arm that issued no release. Arm B has no certificate tool and cannot cite certified evidence by construction, so the study estimates the effect of the added tools as a bundle and leaves the separate contribution of certification unresolved.

Amendment A1 changed the run harness after 29 of 234 jobs; arms B and C each ran ten jobs on the original harness and 68 on the amended one. On the 48 specifications whose B and C sessions shared a harness, the difference was 0.458 (95\% confidence interval 0.323--0.594), and arm B scored higher on none of them.

Two amendments, both made before any confirmatory outcome was read or scored, cut the planned additional repeats, development runs, second provider and oracle-specification arm, and all 234 retained jobs were scored (Table~\ref{tab:identity}). Version 0.2.0 requires refusal evidence that would have rejected one of 26 arm-C refusals in the frozen study. These results characterize one repeat on the frozen implementation, whose checker used the original gap guard; under the released rules the witness certificates of three briefs are undetermined, and the experiment was not rescored.

\end{document}